\documentclass[letterpaper,twocolumn,10pt]{article}
\usepackage{usenix-2020-09}

\usepackage{tikz}
\usepackage{amsmath}
\usepackage{graphicx}
\usepackage{textcomp}
\usepackage{xcolor}
\usepackage{float} 
\usepackage{subfigure} 
\usepackage{pgfplots}
\usepackage{tikz}
\usepackage{algorithm}
\usepackage{algorithmicx}
\usepackage{algpseudocode}
\usepackage{amsmath}
\usepackage{diagbox}

\usetikzlibrary{arrows,shapes,chains,fit}
\tikzset{
  every fit/.style={
    draw,
    thick,
    dashed,
    gray,
    inner sep=5pt
  }
 }
 \newcommand\ModelAcronym{PatternMonitor}

\begin{document}

\date{}

\title{\Large \bf PatternMonitor: a whole pipeline with a much higher level of automation for guessing Android lock pattern based on videos}

\author{
{\rm Yangde Wang}\\
Shanghai Jiaotong University
\and
{\rm Weidong Qiu}\\
Shanghai Jiaotong University
\and
{\rm Yuming Xie}\\
Shanghai Jiaotong University
\and
{\rm Yan Zha}\\
Shanghai Jiaotong University
} 

\maketitle

\begin{abstract}
Pattern lock is a general technique used to realize identity authentication and access authorization on mobile terminal devices such as Android platform devices, but it is vulnerable to the attack proposed by recent researches that exploit information leaked by users while drawing patterns. However, the existing attacks on pattern lock are environmentally sensitive, and rely heavily on manual work, which constrains the practicability of these attack approaches. To attain a more practical attack, this paper designs the PatternMonitor, a whole pipeline with a much higher level of automation system againsts pattern lock, which extracts the guessed candidate patterns from a video containing pattern drawing: instead of manually cutting the target video and setting thresholds, it first employs recognition models to locate the target phone and keypoints of pattern drawing hand, which enables the gesture can be recognized even when the fingertips are shaded. Then, we extract the frames from the video where the drawing starts and ends. These pre-processed frames are inputs of target tracking model to generate trajectories, and further transformed into possible candidate patterns by performing our designed algorithm. To the best of our knowledge, our work is the first attack system to generate candidate patterns by only relying on hand movement instead of accurate fingertips capture. The experimental results demonstrates that our work is as accurate as previous work, which gives more than 90\% success rate within 20 attempts.
\end{abstract}

\section{Introduction}
\label{sec:introduction}

Psychological research shows that for the human brain, visual information is more convenient to remember and recall than character and number information (i.e. PIN- or text-based password) \cite{Weiss2008PassShapes,Angeli2005IsPicWorth}. Benefit from the convenience provided by pattern lock, it is widely applied on various mobile terminal devices including Android devices to provide identity authentication and access authorization for these devices. A recent survey \cite{Ye2017Cracking} showed that nearly 73\% of respondents choose to set up pattern locks on their mobile devices, and the mainstream payment applications such as Paypal and Alipay also enable users to choose the pattern lock their individual login method. However, due to the openness of the unlocking scenario, pattern lock users confront with various security threats. For the sake of user's security and privacy, it is necessary to reveal the 
vulnerability that pattern lock may encounter in practical scenarios.

There are multiple attack systems have been proposed in recent years to exploit the potential vulnerabilities of pattern lock. In 2010, Adam et al. \cite{Aviv2010Sumdge} proposed the Smudge attack by analyzing the leveraged oily residues left on screen to guess unlock patterns. Zhang et al. \cite{Zhang2016Privacy} designed an attack system which reconstructs unlock pattern by monitoring wireless signals. Zhou et al. \cite{Zhou2018PatternListener} developed an app, which collects acoustic signals while victims drawing lock pattern, then forwards these signals to the remote server to recover the unlock pattern. However, the practicality of above attack systems is confronted with restrictions in the real world scenario. Specifically, Smudge attack may be interfered by oily residues from the user's historical operations (not necessarily from the most recent). The Wireless-signal attack requires a complicated network setup, and the success rate may be easily disturbed by environments (e.g.people walking by). The acoustic attack requires that a specified app should be installed in the target device, which is unfeasible in reality. Alternatively, considering each unlock pattern can be regarded as the combination of finite numbers, thus it is possible to guess unlock patterns from video footage. In 2017, Ye et al.\cite{Ye2017Cracking} cracked pattern lock by tracking the victim's fingertips motion from the video footage, but their work heavily relies on continuous fingertips motion, which indicates that the attack may failed if the fingertips are not continuously captured by the camera. Moreover, their attack requires significant manual efforts, including cutting videos, locating fingertips, deciding start typing and end typing moment and etc, this implies that their proposed attack system is hard to be implemented in practical scenarios.

To achieve a more practical attack on the pattern lock, this paper guesses the unlock pattern from the video footage with the recognition model and our designed algorithm, instead of manually locating the fingertips of people who using the target device, in which way we have built the fully automatic attack system PatternMonitor. The advantage of PatternMonitor  includes: First, as a fully automatic attack system,  PatternMonitor can prevent the errors and uncertainties caused by manual operation. Secondly, in comparison with the manual processing in related literatures, the fully automatic mechanism we designed improves the performance of PatternMonitor for processing a video footage. According to the experimental implementation, we can test a video containing the whole procedure of unlocking device including taking out the device, waking up the pattern grid, drawing pattern, and making sure the device is unlocked, within 60 seconds, this is more time-efficient than the manual operations. Thirdly, we provide a novel threaten model against the lock pattern. The experimental results show that in addition to fingertips, it is promising to reconstruct the unlock pattern by capturing other keypoints on the hand. This implies that our work would arouse the reappraising of the academic and industrial communitees to the security of the pattern lock, even though it has been extensively applied worldwide.


\textbf{Contributions.} The contributions of this paper is enumerated as follows.

\textbf{1}. We designed a whole attacking pipeline with a much higher level of automation. 

\textbf{2}. We used multi-point tracking of hands to improve accuracy of recovering use input.

\textbf{3}. We conducted initial experiments under two different recording conditions and showed that our proposed attacking pipeline outperformed the state-of-the-art work.




The remainder of the paper is organised as follows: In Section 2 we briefly overview the related works. Section 3 describes the threat model, and the attack scenarios in detail. Section 4 elaborates the experimental implementation of our proposed automatic attack system, and in Section 5, we give the evaluation of the experimental results. Section 6 discusses the limitation factors of our proposed attack system, and the feasible approach to resist the proposed attack on the pattern lock. The last section concludes this paper.


\section{Related Works.}
\label{sec:related_work}
Mobile devices have accepted widespread popularity in recent years. However, they often serve in privacy sensitive environments, and some of the installed applications \cite{B2011Falling,Brown2014100} which involve sensitive information are also easily exposed to the accessing by unauthorized users.

\textbf{Threat scenarios}
In a nutshell, mobile devices are confronted with the following typical threat scenarios.

\textbf{1}. Adversary taping: This threat scenario requires an adversary to obtain sensitive information from the victim without the victim being aware of it. Prior work \cite{Shukla2019Stealing} demonstrates that with a long distance(e.g. 5m/8m), an adversary can also collects enough information to implement the cracking procedure. In addition,  the increasing types of wearable device \cite{Wang2017Personal,Guerar2019Securing,Lu2018Snoopy} make this attack more convenient and unperceivable.

\textbf{2}. Surveillance camera monitoring: Surveillance cameras have been widely deployed in public places, and the threat of surveillance cameras to personal privacy has been discussed in \cite{Schiff2007Respectful,Winkler2012User}.

\textbf{3}. Shoulder surfing: \cite{Eiband2017Understanding,Rajanna2017A} claim that shoulder surfing attacks are more likely to be conducted in a close distance. however, these researches mainly concentrate on the peep to the static information, such as instant message, web pages and etc., while the dynamic objectives including password inputting and pattern lock drawing are out of consideration (3 unlock patterns and 4 passwords are reported out of 189 samples). Recent works \cite{Ali2014Protecting, Alt2015GravitySpot, Zhou2015Somebody} have proposed some preventive measures against the shoulder surfing, including turning down the screen luminance or changing the operation interface. This paper just sketchily discusses shoulder surfing, since the movement of fingertip is visible in this scenario.

\textbf{Cracking lock pattern}
This kind of attack can be implemented in various ways. In 2010, Aviv et al.\cite{Aviv2010Sumdge} tried to crack lock pattern by collecting information from oily residues left on the screen. The feasibility of this attack approach is questionable sinceusers may wipe away the oily residues for their frequently operations on the mobile phone. Besides, this attack approach can only be succuessfully implemented, on the premise of obtaining the target device. In 2016, Zhang et al. \cite{Zhang2016Privacy} tried to reconstruct lock pattern by monitoring the differences of WIFI signal during drawing lock pattern. This approach is not practical, since the attacker is required to access the router the target device connected before, then conduct the complicated configuration on it. What's more, this attack approach is proved to be environmentally senstive, that is, the success rate of attack is heavily influenced by the  circumanstance change. In 2017, Zhou et al. \cite{Zhou2018PatternListener} came up with a novel approach to reconstruct lock pattern by hearing acoustic signals. However, to hear the acoustic signals, the attacker is required to install a specific APP in the target device, thus this attack approach is also impractical. Subsequently, Abdelrahman et al. \cite{Abdelrahman2017Stay} designed a new attack system which monitors the thermal information during identity authentication. This attack approach is similar to the Smudge attack, but it may be easily disrupted by extra on-screen operations.

In 2018, Ye et al. \cite{ye2018video} tried to crack lock pattern from the video footage that includes the victim's fingertips motion of unlock pattern. This is a very closely related work to this paper.However, our work differs from \cite{ye2018video} in the following aspects:
1.\cite{ye2018video} includes much manual work , such as cutting movies, identify fingertips, setting phone angles etc.Our work uses device recognition module, hand recognition module, Key points tracking module, Trajectory processing module to automatically do these jobs, which means more effective, and less ad hoc.
2.\cite{ye2018video} works only when the fingertips can be seen during the whole drawing pattern procedure, while our work gives a prior list of hand key points including fingertips.Using this list, our tracking module can record each key point's motion individually in descend order, and our pattern recognizing module will generate guess candidates by the optimized trajectory.
3.\cite{ye2018video} make some assumptions (e.g. before or after unlocking, users tend to pause for a few seconds) to complete their attack. Our work gives a more scientific way to prove it, and builds a recognizer based on machine learning algorithm to automatically locate the start and end frame of drawing lock pattern.

\textbf{Video-based attack}
A large number of video-based attack approaches have been developed to break the identity authentication mechanism on the mobile phone. The attack models presented in \cite{Xu2013Seeing,Chen2018Smartphone,Balagani2018SILKTV,balagani2019pilot} can be successfully implemented on the premise of capturing the screen display while typing the text-based key. Therefore, for the user who intentionally shelters the screen while typing, these attack approaches would not work. Maggi et al. \cite{Maggi2011A} published a feasible attack model, which records the video while user typing, and further extracts the privacy information from the recorded video. To attain this attack, their system first benefits from the display feedback mechanism (the enlarged key display while it is being typed), and the camera is required to directly point to the screen of the target device. A similar attack on text-form input including the password has been proposed recently by Yue et al. \cite{Yue2014Blind, Yue2015Blind}. Their attack is achieved by the advanced camera device such as Google glasses to remotely record the screen and fingertip movements, which significantly increases the stealthiness of the attack. The above work aims to the text-form passwords and PIN codes. However, as an extremely common identity authentication mechanism, the security vulnerability of pattern lock is rarely considered.

\section{Background and Threat Model}
\label{sec:background}

\subsection{Pattern Lock}
\label{subsec:pattern_lock}

\begin{figure}
	\centering
	\begin{minipage}[b]{0.5\textwidth}
	\subfigure[Pattern Layout and Pattern 1-2-3-4-5-6-7-8-9]{
		\begin{tikzpicture}
        \node[anchor=south west,inner sep=0] (image) at (0,0) {\includegraphics[width=0.45\textwidth]{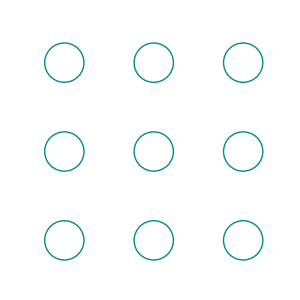}};
    \begin{scope}[x={(image.south east)},y={(image.north west)}]
        \node[] at (0.2,0.9) {1};
        \node[] at (0.2,0.6) {4};
        \node[] at (0.2,0.3) {7};
        
        \node[] at (0.5,0.9) {2};
        \node[] at (0.5,0.6) {5};
        \node[] at (0.5,0.3) {8};
        
        \node[] at (0.8,0.9) {3};
        \node[] at (0.8,0.6) {6};
        \node[] at (0.8,0.3) {9};
    \end{scope}
        \end{tikzpicture}
		\includegraphics[width=0.45\textwidth]{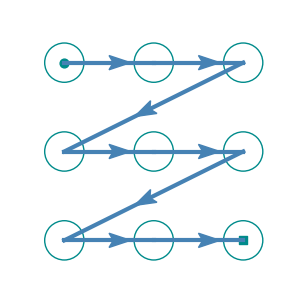}
		\label{fig:pattern_layout}	}
	\subfigure[Valid Pattern Examples]{
			\includegraphics[width=0.22\textwidth]{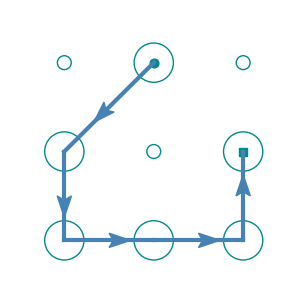} 
			\includegraphics[width=0.22\textwidth]{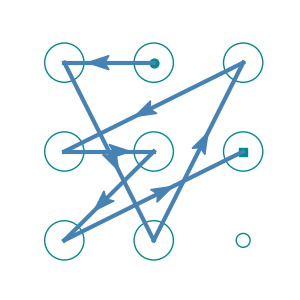}
		\label{fig:valid_patterns}
	}
    \subfigure[Invalid Pattern Examples]{
  		 	\includegraphics[width=0.22\textwidth]{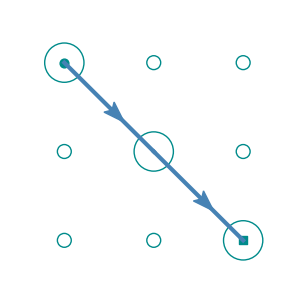}
		 	\includegraphics[width=0.22\textwidth]{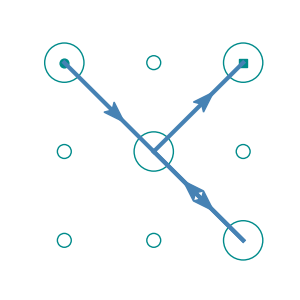}
		\label{fig:invalid_patterns}
    }
	\caption{Some valid and invalid pattern examples.}
	\label{fig:pattern_examples}
	\end{minipage}
\end{figure}
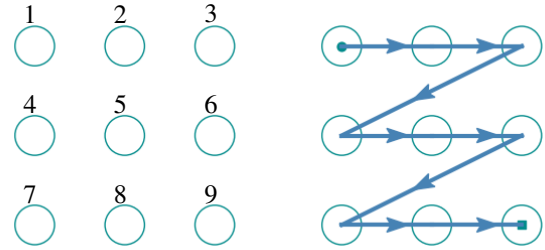
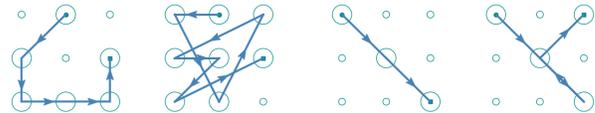

Pattern lock is a popular authentication mechanism which is widely used in mobile devices, both on unlock screen and some APPs such as Paypal and Alipay, since it is easily to remember and recall than the text-form password \cite{Angeli2005IsPicWorth}. Generally, pattern lock requires user to configure a graphic pattern that connects a sequence of contact points arranged in a 3 $\times$ 3 grid, and check the input pattern when someone wants to access the system every time. Figure.~ \ref{fig:pattern_layout} shows the layout, and how we mark these points, while \ref{} gives a specific instance of the graphic pattern. In practice, an valid graphic pattern should satisfy the following requirements \cite{Aviv2015PatternGrid}. 1) It must contains at least 4 points, 2) Each point just for once use, 3) It must be entered without lifting and 4) It may not avoid a previously unselected contact point. With these requirements, in a 3 $\times$ 3 grid, we have 389,112 valid patterns in total. Figure.~\ref{fig:valid_patterns}  and Figure.~\ref{fig:invalid_patterns}  respectively enumerate the valid patterns and invalid ones.

\subsection{Threat Model}
\label{subsec:threat_model}

According to a recent research\cite{2014Studying}, nearly 40$\%$ of the users set the pattern lock rather than the PIN as their identity authentication in their mobile devices. Since the camera on personal mobile devices such as phones and pads have more powerful ability to take photo and video. In addition, for the sake of public safety, surveillance cameras have been increasingly deployed in public place  \cite{barrett2013one,doyle2013eyes}, which also threaten the security of the pattern lock. Naturally, here we describe two typical threat scenarios against the security of pattern lock which are rarely mentioned in previous researches.

\subsubsection{Scenario 1 : Surveillance camera monitoring}
\label{subsubsec:scenario_cameras}

As \cite{doyle2013eyes} says, the number of surveillance cameras are growing all over the world. These cameras are deployed not only in public place, but also in domestic such as coffee houses, hotels, and malls. In these places, surveillance cameras are usually set up near the ceiling, that is, about 3 metres above the floor. We choose this scenario as a typical threaten for two reasons: 1). Surveillance cameras set up in these places are easy to be invaded \cite{costin2016security}. Due to employees of these places are always lacking of knowledge of security, hackers may take advantage of this fact, then hacks into the cameras to steal the video stream. 2). The user in these places may not be aware of being monitored by surveillance cameras, which will lead them to let their guard down to unlock their devices in an easy way, i.e. just draw the unlock pattern without any cover. An example of this scenario is illustrated as Figure~\ref{fig:scenario_surveillance}.

\subsubsection{Scenario 2 : Face-to-face taping} 
\label{subsubsec:scenario_face_to_face}

\begin{figure}[htbp]
    \centering
    \subfigure[Scenario 1 : Surveillance camera monitoring]{
    \includegraphics[width=10em]{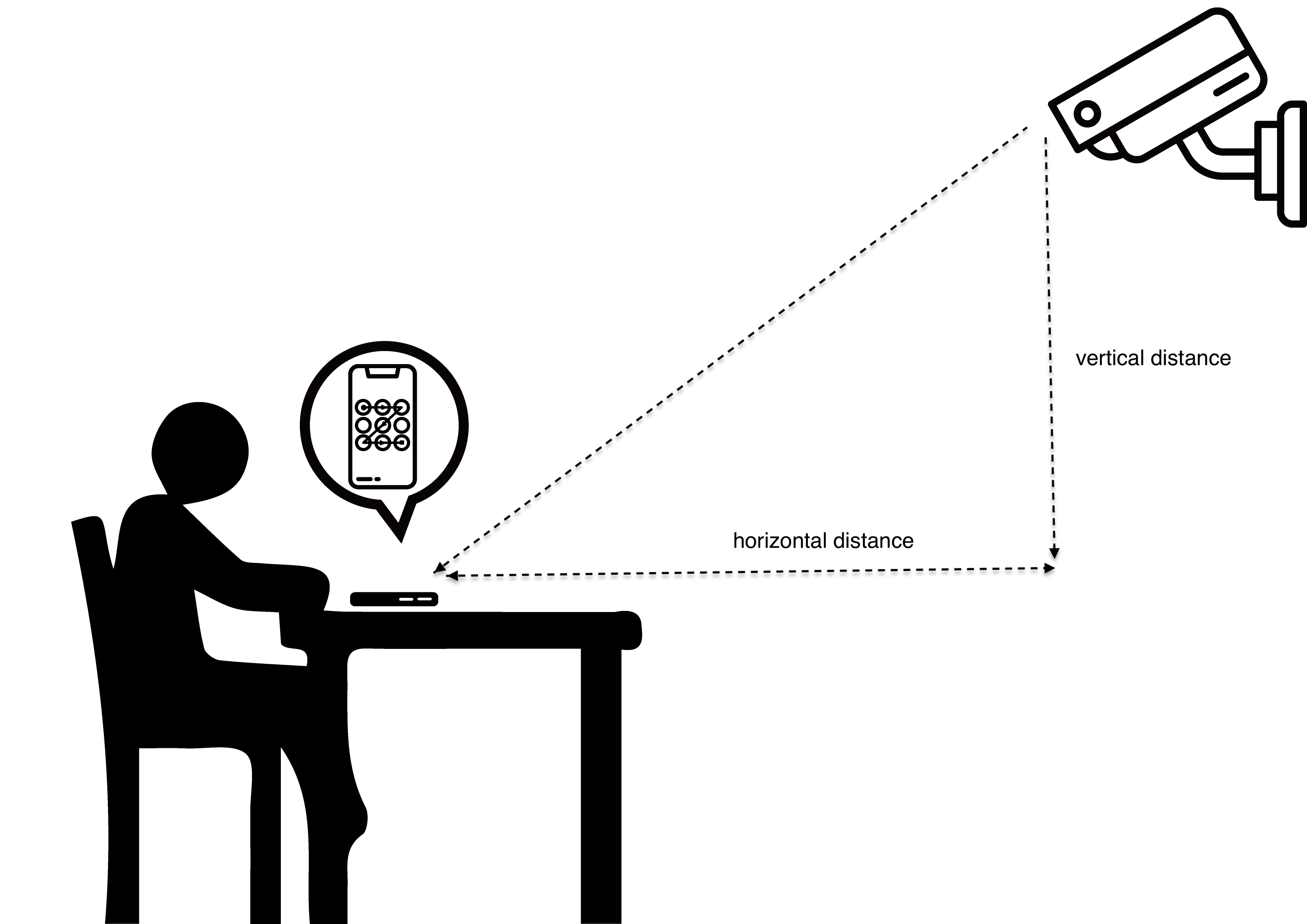}
    \label{fig:scenario_surveillance}
    }
    \quad
    \subfigure[Scenario 2 : Face-to-face taping]{
    \includegraphics[width=10em]{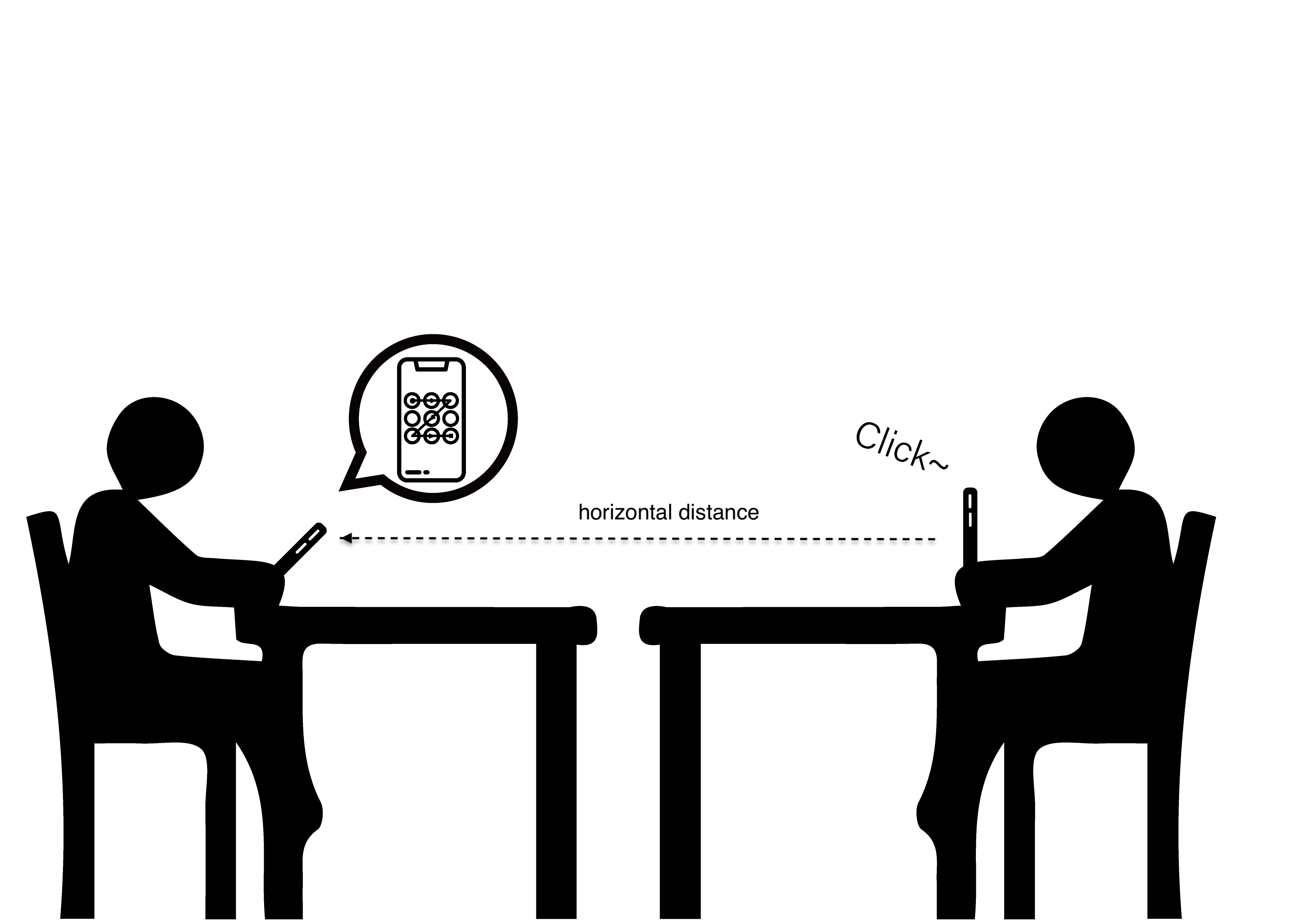}
    \label{fig:scenario_face_to_face}
 	}
	\caption{Examples of attack scenarios discussed in our paper.}
\end{figure}

The biggest difference between this scenario and scenatio~\ref{subsubsec:scenario_cameras} is that the screen of target device and user's fingertips of his pattern-drawing hand are all inevitably missing. To the best of our knowledge, ours is the first work that can generate pattern guess candidates only on the basis of observed hand movements. Figure~\ref{fig:scenario_face_to_face} gives an example of this scenario. It shows that, when the victim draws a pattern, an adversary may sit not very far away from him, taping the victim's actions. In this scenario, the victim usually has an illusion of self-security for following reasons: 1). He/She may believe that the screen of his/her phone and fingers are not visible. 2). The adversary can conceal his actions in divers ways, such as using small wearable devices to tape, or pretending to watch movies but actually taping with mobile phones.  

\subsection{New Challenges}
\label{subsec:new_challenges}

To overcome the flaws prior work has which make the attack inapplicable, we need to address some new challenges:

\textbf{\emph{Challenge 1:}} TBD

\textbf{\emph{Challenge 2:}} TBD

\textbf{\emph{Challenge 3:}} Which point on the hand can be considered to be key point? Will different key points affect the accuracy of tracking hand motion? Especially on scenario 2, since we can not see the fingertips, how to track hand motion?

\textbf{\emph{Challenge 4:}} How to locate the frames that the user starts to drawing the pattern, and the ones that user finishes drawing?

\textbf{\emph{Challenge 5:}} Under both scenarios, no matter surveillance cameras or hand holding devices, have vertical distance off the target device, so we can not simply use affine transformation to transform the trajectory to the pattern of user's perspective. Then how can we generate guess candidates from the trajectories?

\subsection{Assumptions}
\label{subsec:assumptions}

To illustrate how our \ModelAcronym works, we make several reasonable assumptions, which can clarify the attack scenario, but not simplify it, as following:

\emph{Assumption 1:} User use pattern lock as the authentication method, which means, if he wants to unlock screen, he will do the following actions in sequence: (1). take out the target device (i.e. mobile phone). (2). awake the device. (3). start drawing the pattern.

\emph{Assumption 2:} The layout of gird on target device is 3 $\times$ 3, for most modern mobile phones use this layout and do not offer other alternative options (e.g. 4 $\times$ 4, or 6 $\times$ 6 grid etc.)

\emph{Assumption 3:} The video footage should contain parts of the drawing hand, but fingertips, the console’s geometry, or any content displayed on the screen are not necessary. We consider this assumption is reasonable for people may cover the drawing fingertip but not all drawing hand. However, lacking of visual of fingertips, knowledge of console’s geometry, displayed content left on the screen, will make prior video-based work focusing on pattern lock\cite{Ye2017Cracking, ye2018video, Shukla2014beware, Aviv2010Sumdge} out of action.

\emph{Assumption 4:} We assume that user draw the correct pattern. Although it is possible that user may draw wrong pattern by mistake, but we mainly focus on how to automatically guess the correct pattern from user's motion, so the incorrect motion is not under our consideration. Furthermore, we also ignore the situation that user just take out his device but not unlock it (e.g. sweep dust on the screen).

\emph{Assumption 5:} We assume that the device's head has the same direction of user's face. With the technique of head pose estimation improving\cite{murphy2008head,  Ruiz_2018_CVPR_Workshops}, we can easily get the direction of user's head, so we use this direction as the target device's orientation.

\emph{Assumption 6:} For the attack scenario mentioned in \ref{subsubsec:scenario_cameras}, we only consider surveillance cameras that deployed in relative small indoor space, such as coffee houses, or malls, but not in public places, or big indoor space such as railway stations or airports. We have this assumption because we do not have professional cameras these places have which can record vivid videos even they are deployed in very high place (e.g. 5 metres or higher). In other words, we only consider surveillance cameras which are put up with height of near 3 metres.

\section{Attack Detail}
\label{sec:attack_detail}

In this section, we first sketchily describe the workflow of the \ModelAcronym, then elaborate the detailed implementation steps of our attack system.

\subsection{Overview}
\label{subsec:overview}

\begin{figure*}[htbp]
    \centerline{
    \includegraphics[width=52em]{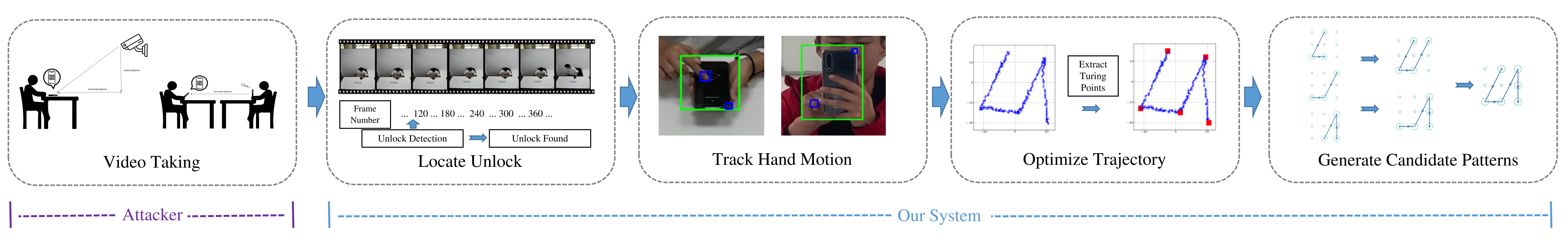}
	}
		\caption{ The workflow of attack. An adversary takes a video of a user's pattern input, inputs it into the \ModelAcronym, and then gets the possible patterns. 
		}
\label{fig:workflow}
\end{figure*}

Figure.~\ref{fig:workflow} shows the workflow of our system. Videos that contain an unlocking process filmed by surveillance cameras or face-to-face shooting are input into the system. To analyse the video footage, our system uses the following steps to automatically generate candidate patterns. 

\textbf{Locate start point of drawing unlock pattern.} The input video usually does not perfectly cover just the scene in which the victim draw the graphic pattern, that is, it also contains the irrelevant frames. In this step, we extract the frames that the user starts to draw pattern by the following substeps: 1). Using object detection algorithm to identify target phone. 2). Magnifying the area of the bounding box of target phone, and start to detect user's drawing hand. 3). Determining whether the user is operating the mobile phone by the relative position between phone and key points of hand. Examples of these substeps are shown in Figure.~\ref{fig:find_start_3}, \ref{fig:find_start_4}, \ref{fig:find_start_5}. After these three substeps, the approximate starting position of drawing unlock pattern in the video can be considered to be found.

\textbf{Track hand motion.} Once the start of drawing pattern is found and key points of hand are detected, tracking algorithms can be employed to locate key points in each successive frame. Using the relative position between the phone and the hand, a trajectory which reflects the user's pattern drawing is produced.

\textbf{Optimize trajectory.} After generating the trajectory of hand motion, We extract the turning points from the trajectory in this step to make identifying the pattern lock more efficient. The original trajectory can be simulated by a few turning points, and the turning points correspond to the ciphers. For example, the trajectory in Figure.~\ref{fig:dp_example} can be simulated by 7 points.

\textbf{Generate candidate patterns.} In this step, the processed trajectory (turning points) is mapped to possible patterns. We divide the trajectory into small parts, each part is mapped to ciphers. By combining these small parts, we obtain all possible results. The candidate patterns are sorted according to their confidence and are tried one by one on target device.

\subsection{Locate Unlocking}
\label{subsec:locate_unlocking}

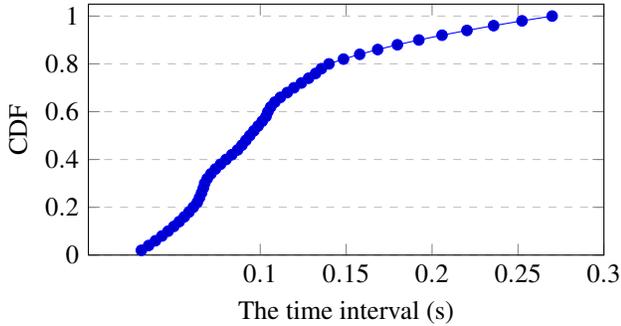
\begin{figure}[htbp]
    \centerline{

\pgfplotsset{width=24em,height = 14em}
\pgfplotstableread[col sep=comma]{data/StartTimeCDF.csv}\datacsv
\begin{tikzpicture}
\begin{axis}[
    xlabel={The time interval (s)},
    ylabel={CDF},
    xmin=0, xmax=0.30,
    ymin=0, ymax=1.05,
    xtick={0.1,0.15,0.2,0.25,0.3},
    ytick={0,0.2,0.4,0.6,0.8,1},
    ylabel near ticks,
    xlabel near ticks,
    legend pos=north west,
    ymajorgrids=true,
    grid style=dashed,
]

\addplot table [x=time, y=CDF, col sep=comma,no markers]{\datacsv};
\end{axis}
\end{tikzpicture}
	}
		\caption{The cumulative distribution function(CDF) of the time interval between pattern drawing and other on-screen activities.
		}
\label{fig:startTimeCDF}
\end{figure}

The first big challenge \ModelAcronym{} faces is how to automatically locate frames where target user starts to draw unlock pattern. Previous work of video-based attacks on pattern \cite{Ye2017Cracking, Xu2013Seeing} find the video segment based on the hypothesis that the user's fingertip often pauses for nearly 1.5 seconds before and after unlocking. However, as shown in Figure.~\ref{fig:startTimeCDF}, based on our analysis of our own data sets, although the drawing habits are vary for different users, there is no feature of 1.5 seconds' pause to judge whether the user starts or ends drawing. In the following, we give a new method to locate the frames that target user start are about to drawing patterns. Unlike the previous works, our attack system only give a rough start point, but it exhibits the excellent experimental results.

\begin{figure*}[htbp]
    \centering
        \subfigure[Process the video stream in real time to judge whether the mobile phone appears in the video ]{
        \includegraphics[width=7em]{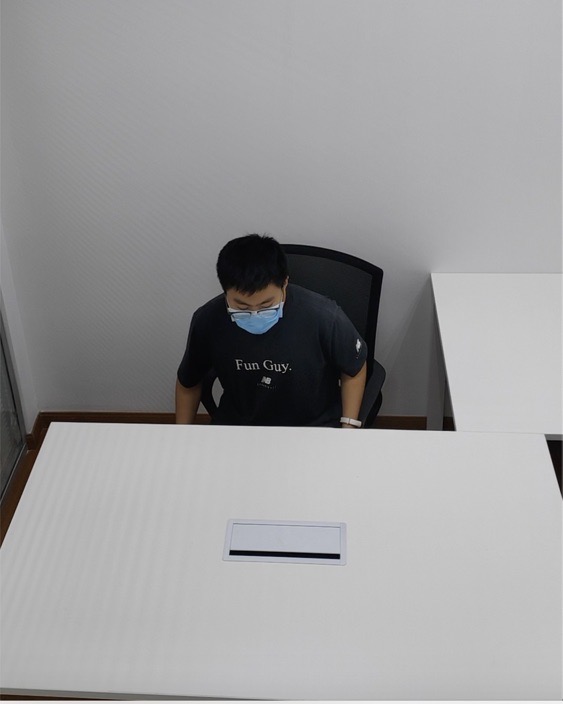}
        \label{fig:find_start_1}}
        \hspace{1em}
        \subfigure[The Mobile phone detected and the system return it's bounding box]{
        \includegraphics[width=7em]{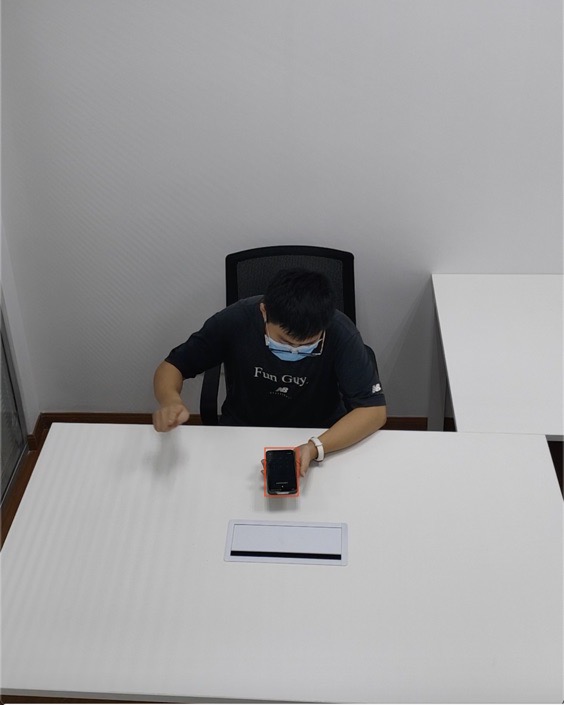}
        \label{fig:find_start_2}}
        \hspace{1em}
        \subfigure[Enlarge bounding box as detection area]{
        \includegraphics[width=7em]{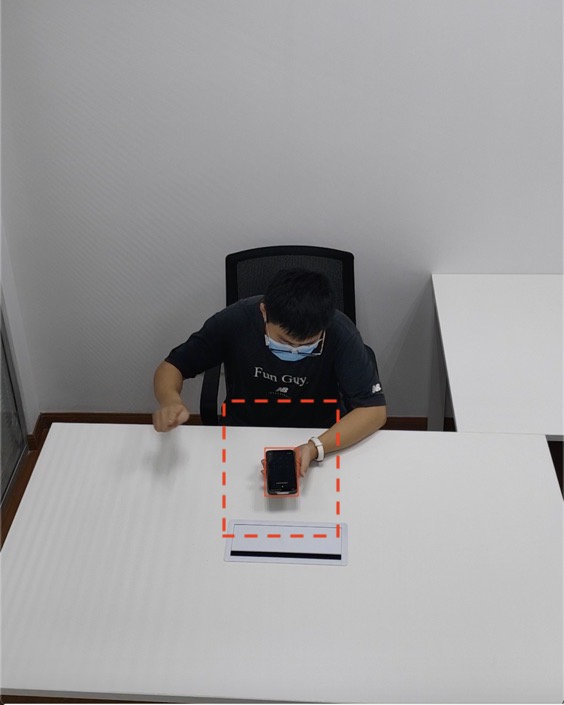}
        \label{fig:find_start_3}}
        \hspace{1em}
        \subfigure[Detect the number of hand keypoints in detection area]{
        \includegraphics[width=7em]{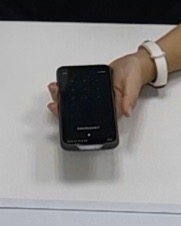}
        \label{fig:find_start_4}}
        \hspace{1em}
        \subfigure[The start of unlocking located as the number of hand keypoints in detection area is greater than threshold]{
        \includegraphics[width=7em]{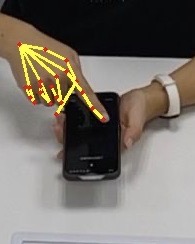}
        \label{fig:find_start_5}}
	\caption{Find Start workflow}
	\label{fig:find start workflow}
\end{figure*}
 

In both scenarios, we apply an object detection algorithm to detect phones in each frame of the video. One example of the workflow of the attack to Scenario. 1 is shown in Figure.~\ref{fig:find start workflow}. Once the target phone is detected, we magnify the bounding box of the phone as hand-detection area by using OpenPose \cite{Simon_2017_CVPR} to identify keypoints of user's hand only in this scope. If \ModelAcronym{} finds that the keypoints we predefined appears in the hand-detection area, we consider the user is about to draw unlocking pattern, and these frames will be labelled as starting unlocking. More details on implementation are described as follows.

\paragraph{Detect phone and phone corner} To identify phones appear in videos, we introduce YOLOv3\cite{yolov3}, a SOTA, real-time object detection system to get this job done. First, we took 1,000 pictures of mobile phones with different models as the phone training set.We also make another 500 pictures of mobiles and manually label phone corner in them to be another part of phone training set, for the reason that we also need automatically mark phones' corner as the reference for keypoints of drawing hand to generate trajectories. Second, to improve the capability for generalization of YOLOv3, we applied 20 types of image augmentation methods to the original phone training set of 1,500 images, which led us to have 30,000 images in total to train YOLOv3, 20,000 images only have phones and 10,000 images have both phones and corners respectively.

\begin{figure}[htbp]
    \centering
    \subfigure[The 21 hand keypoints in HandPose]{
        \includegraphics[width=6em, height=8em]{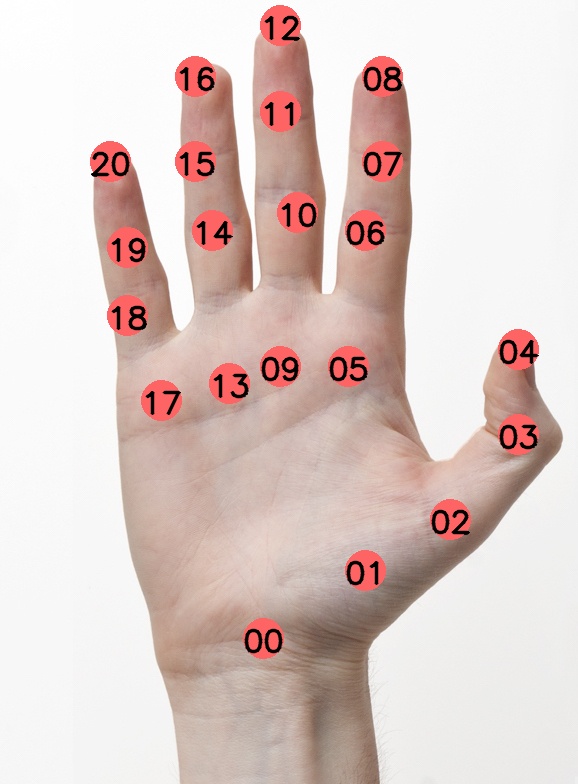}
        \label{fig:handpose_keypoints}
    }
    \subfigure[The prob map of 06th, 07th and 08th keypoint used in Scenario.~\ref{subsubsec:scenario_cameras}]{
        \includegraphics[width=8em,
        height=8em]{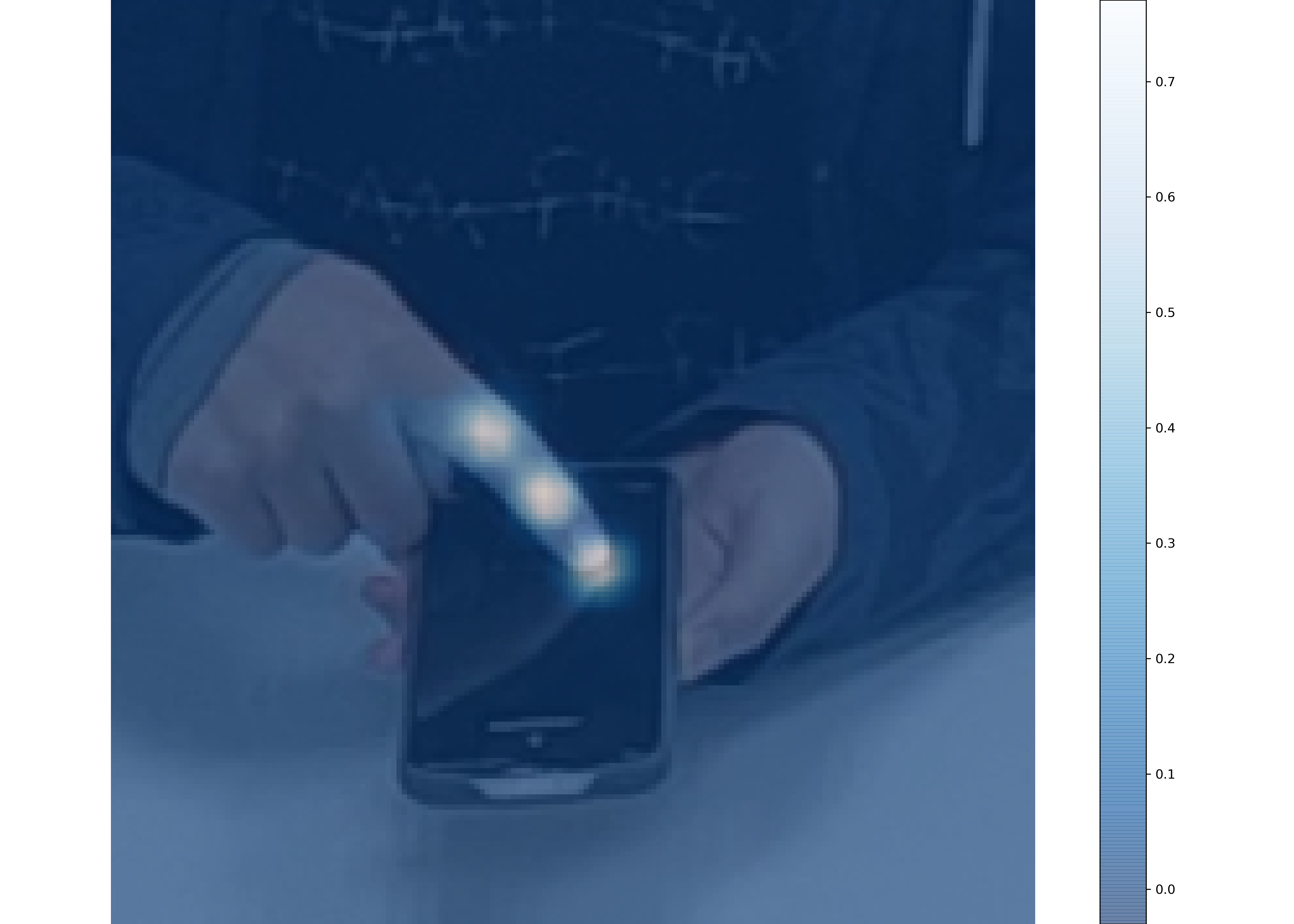}
        \label{fig:handpose_s1}
    }
    \subfigure[The prob map of 17th, 18th and 19th keypoint used in Scenario.~\ref{subsubsec:scenario_face_to_face}. ]{
        \includegraphics[width=8em,
        height=8em]{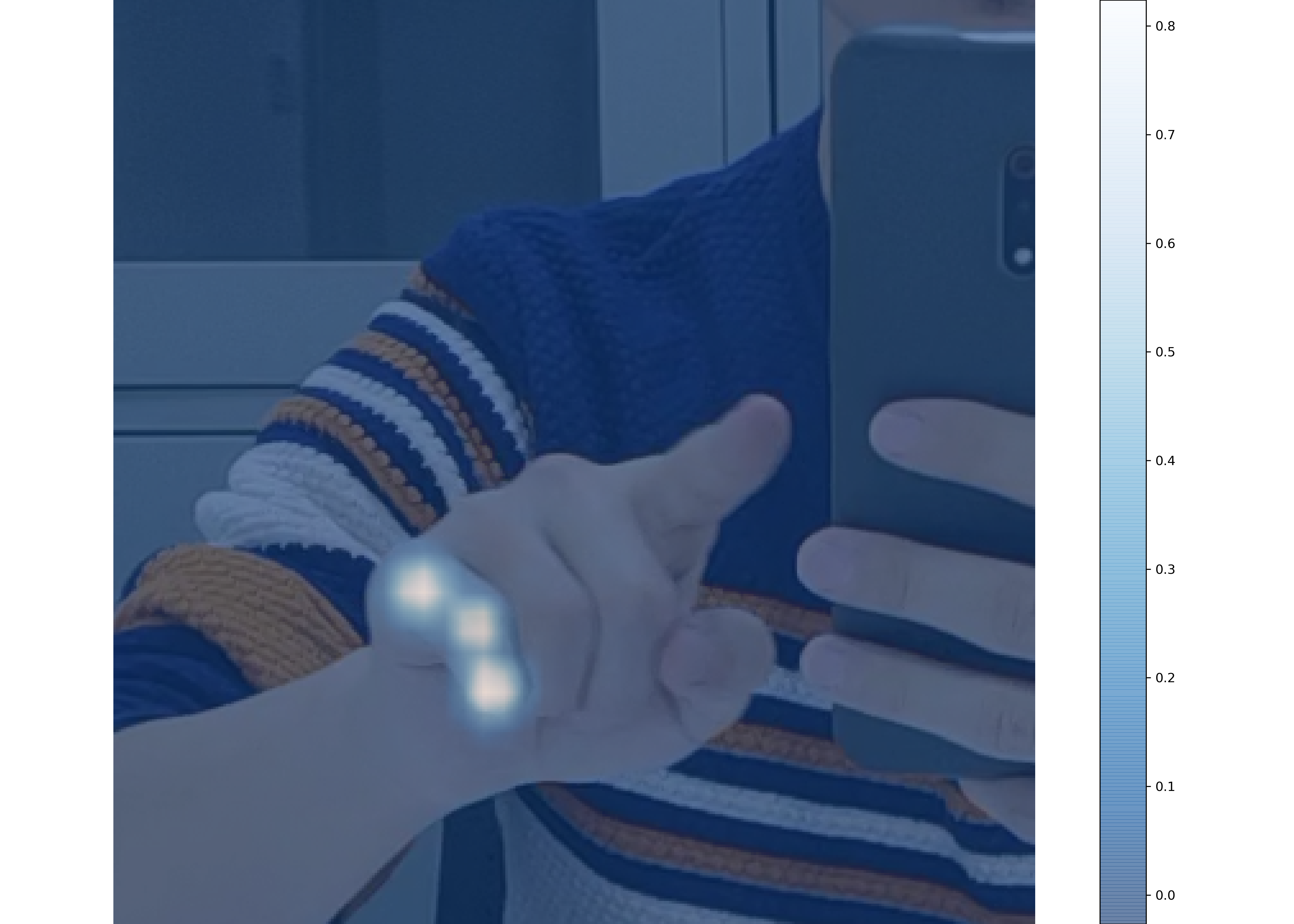}
        \label{fig:handpose_s2}
    }
	\caption{The example of finger keypoints}
	\label{fig:handpose}
\end{figure}

\paragraph{Detect keypoints of drawing hand} OpenPose \cite{Simon_2017_CVPR} is a mature artificial intelligence model which uses a multi-camera system to train fine-grained detectors for keypoints that are prone to occlusion, such as the joints of a hand. It is similar to finding keypoints on Face Detection or Body Estimation\cite{Wei2016ConvolutionalPM,Simon_2017_CVPR}, but different from Hand Detection since in that case, they treat the whole hand as one object. The model produces 22 keypoints. The hand has 21 points while the 22nd point signifies the background. The points are as shown in Figure.~\ref{fig:handpose_keypoints}. 

If there is only one hand appears in detection area, as Figure.~\ref{fig:handpose_s1} shows, the location of each keypoint can be easily detected. However, in pratical, as Figure.~\ref{fig:handpose_s2}shows, users tend to use one hand to hold the phone and another hand to draw unlocking pattern. In that case, OpenPose will give multiple interest areas for it isn't able to distinguish between right hand and left hand. If that happens, we just return all interest areas of keypoints and use the algorithm Section.~\ref{subsec:track_hand_motion} describes to exclude the wrong areas by collecting tracking information.

\subsection{Track Hand Motion}
\label{subsec:track_hand_motion}

In this model, \ModelAcronym{} takes a set of frames that the user is about to draw unlocking pattern with phone, corner of this phone, keypoints of drawing hand labeled in each of them as input, and generate trajectories of validated keypoints relative to corner of target phone by using an introduced tracking algorithm named CSRT\cite{Lukezic2017csrt}.


\paragraph{Selection of keypoints on drawing hand} One of the big differences between Scenario. 1 and Scenario. 2 is the visual parts of drawing hand. In Scenario. 1, usually all parts of hand are visible throughout the unlocking process, while in Scenario. 2, only parts of hand can be seen. Under this situation, we select the keypoints labeled as number 6, 7, and 8 to be alternatives in Scenario. 1, and number 18, 19, 20 to be alternatives in Scenario. 2. Attackers can also choose other keypoints based on what video they get, and we use this configuration only to show how \ModelAcronym works. In Section.~\ref{subsec:generate_candiate_patterns}, we give a way of how to combine the trajectories to generate more reliable candidate patterns.




\paragraph{Selection of object tracking algorithm} Many object tracking algorithms have been proposed such as TLD, BOOSTING, CSRT etc.. Previous work\cite{Ye2017Cracking, Shukla2014beware,Shukla2019Stealing} all used TLD as their tracing tool, but we decide to use CSRT instead for the following three reasons: 1). To make our system completely automatic, no manual work should be involved in tracking. However, TLD need manually check when the tracking results are in low confidence\cite{Shukla2014beware}. 2). Most pattern drawing process last no more than 10 seconds, so our work can be seen as a short-term tracking problem. In terms of that situation, we choose CSRT because it is designed for short-term tracking while TLD is for long-term tracking. 3). Our work has high requirements for the precision of object tracking, and on the OTB100 benchmark, CSRT scores highest in average precision plot\cite{Lukezic2017csrt}.

\paragraph{Generate trajectory and check validation } Given frames with phone, corner of phone, keypoints of hand labeled in them, we use CSRT to track motions of corner of phone, and keypoints of hand, then generate raw trajectory by calculating the relative distance between the two bounding boxes frame by frame. Note that these frames still contain wrong areas which are not belong to user's drawing hand, we can exclude these areas by two assertions: 1). bounding box of wrong areas always moves slightly. 2). bounding box of wrong areas always moves differently from others. Then, we adjust the raw trajectory by excluding frames before the bounding boxes of keypoints of hand enter the phone area and after the boxes leaves. Note that this raw trajectory still includes extra user actions such as sliding the screen to activate the phone or moving outside the phone after drawing the pattern, we will use another algorithm mentioned in \ref{} to deal with it. 

\begin{figure}[ht]
    \centering
	\subfigure[The example of tracking in scenario 1.]{
    \includegraphics[width=10em,height = 10em]{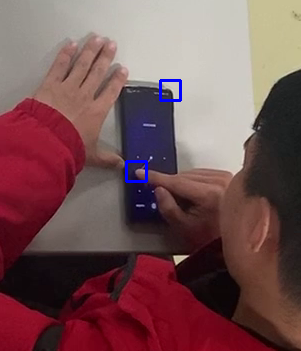}
    }
    \subfigure[The trajectory of example (278369) in scenario 1]{
    \pgfplotsset{width=14em,height = 14em}
    \pgfplotstableread[col sep=comma]{data/trajectory_s1_278369_.csv}\datacsv  
    \begin{tikzpicture}                 
    \begin{axis}[
    xmajorgrids=true,
    ymajorgrids=true,
    xticklabels={},
    yticklabels={},]
    \addplot [blue,quiver={u=\thisrow{U},v=\thisrow{V}},-stealth,]table [x=X, y=Y]{\datacsv};
    \addplot [scatter,only marks,mark = triangle, mark size = 2 pt,
    point meta=1
    ]table [x=X, y=Y,meta = C]{\datacsv};               
    \end{axis}                         
    \end{tikzpicture} 
    }
    \quad
    \subfigure[The example of tracking in scenario 2]{
    \includegraphics[width=10em,height = 10em]{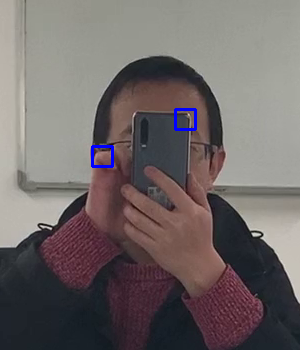}
    }
    \subfigure[The trajectory of example (5298674) in scenario 2]{
    \pgfplotsset{width=14em,height = 14em}
    \pgfplotstableread[col sep=comma]{data/trajectory_s2_5298674_.csv}\datacsv 
    \begin{tikzpicture}                 
    \begin{axis}[
    xmajorgrids=true,
    ymajorgrids=true,
    xticklabels={},
    yticklabels={},]
    \addplot [blue,quiver={u=\thisrow{U},v=\thisrow{V}},-stealth,]table [x=X, y=Y]{\datacsv}; 
    \addplot [scatter,only marks,mark = triangle,mark size = 2 pt,
    point meta=1
    ]table [x=X, y=Y,meta = C]{\datacsv};               
    \end{axis}                         
    \end{tikzpicture} 
    }
\caption{Typical tracking scenarios and the trajectory results. The blue rectangle in pic (a) and pic (c) are bounding boxes. Pic (b) is the trajectory of pic(a) and the pattern is 2-7-3-6-9. Pic (d) is the trajectory of pic(b) and the pattern is 5-2-9-8-6-7-4. }
\label{fig:track_demo}
\end{figure}

However, this tracking process may fail if XXXXXX. To clarify how this model works, we use Figure.~\ref{fig:trackflowchart} to demonstrate the whole process, and Figure~.\ref{fig:track_demo} gives examples of tracking results under the two scenarios we mentioned in Section.~\ref{subsec:threat_model}.

\subsection{Optimize Trajectory}
\label{subsec:optimize_trajectory}

Since raw trajectories always contain noise and redundancy, it is hard to generate candidate patterns directly based on them. In this step, we describe how we optimize trajectories in detail to raise the efficiency of pattern identifying. 

\begin{figure}[t]
    \centerline{
        \pgfplotsset{width=18em,height = 18em}
    \pgfplotstableread[col sep=comma]{data/dp_tra_s1_278369_.csv}\dpdata 
    \pgfplotstableread[col sep=comma]{data/trajectory_s1_278369_.csv}\datacsv 
    \begin{tikzpicture}                 
    \begin{axis}[
    xmajorgrids=true,
    ymajorgrids=true,
    x tick label style={font=\tiny,yshift=0.6ex},
    y tick label style={font=\tiny,xshift=0.6ex},
    ]
    \addplot [blue,quiver={u=\thisrow{U},v=\thisrow{V}},-stealth,]table [x=X, y=Y]{\datacsv};  
    \addplot [scatter,only marks,mark = triangle,
    point meta=explicit symbolic,
    scatter/classes={
    T={mark=square*,red,mark size = 4pt},
    F={mark=triangle,blue,}},
    ]table [x=X, y=Y,meta = C]{\datacsv};               
    \addplot+[nodes near coords,nodes near coords style={anchor=west,xshift=2pt},only marks,mark size =0,point meta=explicit symbolic]table [x=X, y=Y,meta = 0]{\dpdata};  

    \end{axis}                         
    \end{tikzpicture} 
	}
	\caption{Using RDP algorithm to identify the turning points in trajectory. There are 222 points in original trajectory. 7 turning points are identified by RDP algorithm.  }
	\label{fig:dp_example}
\end{figure}
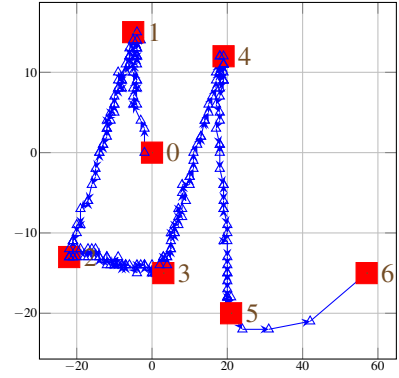

As shown in Figure \ref{fig:dp_example}, one trajectory can be simulated by some turning points. A line segment defined by two turning points is considered as the smallest constituent part of a pattern. Based on these line segments, we can use our algorithm mentioned in Section.~\ref{subsec:generate_candiate_patterns} to generate candidate patterns. Our goal in this step is to extract turning points from the raw trajectory and handle the overlap trajectory in some circumstances. 

We employ the Ramer-Douglas-Peucker\cite{douglas1973RDP} (RDP) algorithm to extract turning points from trajectory. RDP is an algorithm to reduce the points in a curve but keep the shape, so we can use line segments RDP processed to approximate the raw trajectory. Figure.~\ref{fig:dp_example} shows how RDP works: 1). First, we have a start point (labeled as 0), and an end point (labeled as 6), and RDP uses these two points to form a line segment. 2). Calculate all other points in the trajectory, and a new point will be added to form new line segments with existing vertices if this point has the longest distance to the line segment, and the distance of it reaches a predefined threshold. As Figure.~\ref{fig:dp_example} shows, the point labeled as 2 will be considered to be a new turning point, and new line segments 0~2, 2~6 are formed. 3). Repeat substep 2) until no points need to be considered to be turning points, such as Figure.~\ref{fig:dp_example} shows. After RDP algorithm is done, we will get all possible turning points which form the trajectory.


Intuitively, trajectory of pattern who has overlapping line segments is an obstacle to extracting turning points using RDP, but in practical, RDP still works well under this situation. There are two kinds of trajectories who have overlapping: 1). Overlapping happens at start or end or both of trajectory; 2). Overlapping happens in the middle of trajectory. Our algorithm of RDP can deal with the second kind of overlapping trajectory to extract turning points. For the first kind, note that the raw trajectory still includes some redundant frames, i.e. the process of sliding the screen to activate the phone and moving outside the phone after drawing the pattern, so RDP algorithm still can find the turning point with overlapping at start or end. 
Take Figure.~\ref{fig:overlapping_example} for example, point 2 will not be considered as a turning point because its distance to line 0-3 is near zero. We solved this problem by keeping redundant trajectories at the beginning and end.


\subsection{Generate Candidate Patterns}
\label{subsec:generate_candiate_patterns}

\begin{figure}[t]
    \centering
    \begin{tikzpicture}
    \node[anchor=south west,inner sep=0] (image) at (0,0) {\includegraphics[width=0.42\textwidth]{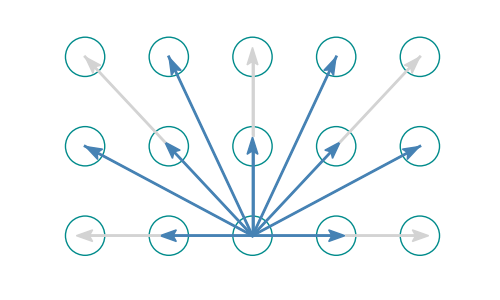}};
    \begin{scope}[x={(image.south east)},y={(image.north west)}]
        \node[] at (0.3,0.3) {A};
        \node[] at (0.2,0.6) {B};
        \node[] at (0.3,0.6) {C};
        \node[] at (0.3,0.9) {D};
        \node[] at (0.5,0.6) {E};
        \node[] at (0.7,0.9) {F};
        \node[] at (0.7,0.6) {G};
        \node[] at (0.8,0.6) {H};
        \node[] at (0.7,0.3) {I};
        \node[] at (0.2,0.3) {A$^{'}$};
        \node[] at (0.2,0.9) {C$^{'}$};
        \node[] at (0.5,0.15) {O};
    \end{scope}
    \end{tikzpicture}
    \caption{All possible angles in patterns.}
    \label{fig:all_possible_angles}
\end{figure}
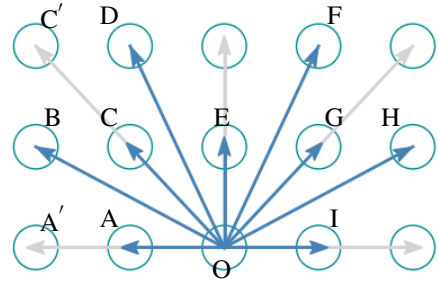

\begin{figure}[!htb]
    \subfigure[The distribution of LSL]{
        \centerline{
    \pgfplotsset{width=24em,height = 14em}
    \pgfplotstableread[col sep=comma]{data/distance_density.csv}\datacsv
    \begin{tikzpicture}
    \begin{axis}[
    smooth,
    xlabel= Distance,
    ylabel= Probability Density,
    ymin = 0,
    xmin = 0,
    ymajorgrids=true,
    ylabel near ticks,
    xlabel near ticks,
    legend entries = {1,$\sqrt{2}$, 2, $\sqrt{5}$, 2$\sqrt{2}$},
    legend style = {
    at = {(0.98, 0.98)}}
    ]
    \addplot[red] table[x=0, y=1] {\datacsv};
    \addplot[blue] table[x=0, y=2] {\datacsv};
    \addplot[green] table[x=0, y=3] {\datacsv};
    \addplot[brown] table[x=0, y=4] {\datacsv};
    \addplot[yellow] table[x=0, y=5] {\datacsv};
    \end{axis}
    \end{tikzpicture}
    }
        \label{fig:distance_dentisity}
    }
    \subfigure[The distribution of angle]{
        \centerline{
    \pgfplotsset{width=24em,height = 14em}
    \pgfplotstableread[col sep=comma]{data/angle_density.csv}\datacsv
    \begin{tikzpicture}
    \begin{axis}[
    smooth,
    xlabel= Angle,
    ylabel= Probability Density,
    ymin = 0,
    xmin = 0,
    xmax = 160,
    x tick label style={font=\small},
    y tick label style={/pgf/number format/fixed},
    ytick={0.05,0.10,0.15,0.20},
    ymajorgrids=true,
    ylabel near ticks,
    xlabel near ticks,
    legend entries = {18$^{\circ}$, 27$^{\circ}$, 37$^{\circ}$, 45$^{\circ}$, 53$^{\circ}$, 63$^{\circ}$, 72$^{\circ}$, 90$^{\circ}$, 117$^{\circ}$, 135$^{\circ}$},
    legend columns = 3,
    legend style = {
    at = {(0.98, 0.98)}}
    ]
    ]
    \addplot[red] table[x=0, y=1] {\datacsv};
    \addplot[blue] table[x=0, y=2] {\datacsv};
    \addplot[yellow] table[x=0, y=3] {\datacsv};
    \addplot[pink] table[x=0, y=4] {\datacsv};
    \addplot[black] table[x=0, y=5] {\datacsv};
    \addplot[gray] table[x=0, y=6] {\datacsv};
    \addplot[orange] table[x=0, y=7] {\datacsv};
    \addplot[purple] table[x=0, y=8] {\datacsv};
    \addplot[green] table[x=0, y=9] {\datacsv};
    \addplot[brown] table[x=0, y=10] {\datacsv};
    \end{axis}
    \end{tikzpicture}
    }
        \label{fig:angle_dentisity}
    }
\caption{The distribution of LSL and angle in trajectories}
\label{fig:dentisity}
\end{figure}
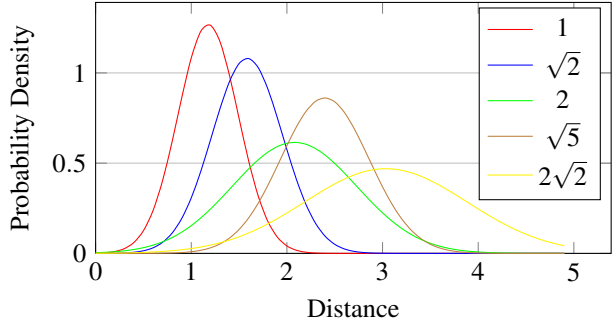
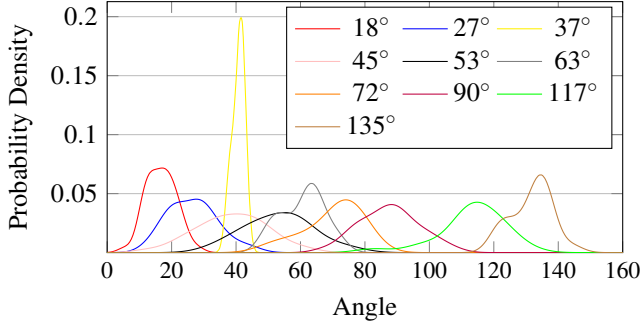

\begin{figure}[htbp]
    \begin{minipage}[b]{0.5\textwidth}
        \subfigure[The standard trajectory of example(278369)]{
            \includegraphics[width=0.4\textwidth]{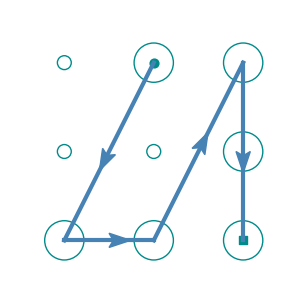}
        }
        \hspace{1em}
        \subfigure[The trajectory of example (278369) in scenario]{
                \pgfplotsset{width=0.5\textwidth,height = 0.5\textwidth}
                \pgfplotstableread[col sep=comma]{data/dp_tra_s1_278369_.csv}\dpdata 
                \pgfplotstableread[col sep=comma]{data/trajectory_s1_278369_.csv}\datacsv 
                \begin{tikzpicture}                 
                \begin{axis}[
                xmajorgrids=true,
                ymajorgrids=true,
                x tick label style={font=\tiny,yshift=0.6ex},
                y tick label style={font=\tiny,xshift=0.6ex},
                ]
                \addplot [blue,quiver={u=\thisrow{U},v=\thisrow{V}},-stealth,]table [x=X, y=Y]{\datacsv};  
                \addplot [red,quiver={u=\thisrow{U},v=\thisrow{V},every arrow/.append style={line width=1pt}},-stealth,]table [x=X, y=Y]{\dpdata};  
                \addplot [scatter,only marks,mark = triangle,
                point meta=explicit symbolic,
                scatter/classes={
                T={mark=square*,red,mark size = 4pt},
                F={mark=triangle,blue,}},
                ]table [x=X, y=Y,meta = C]{\datacsv};               
                \addplot+[nodes near coords,nodes near coords style={anchor=west,xshift=2pt},only marks,mark size =0,point meta=explicit symbolic]table [x=X, y=Y,meta = 0]{\dpdata};  
            
                \end{axis}                         
                \end{tikzpicture} 
            \label{fig:opt_trajectory_example}
        }
    \end{minipage}
	\caption{The example of deformation between standard pattern and trajectory}
	\label{fig:}
\end{figure}

In this step, the trajectory optimized will be mapped to possible lock patterns. Our approach generates as many candidate patterns as possible and sorts them by its confidence in descending order. \cite{Ye2017Cracking} used the geometry information of the whole trajectory to get most-likely patterns. But for \ModelAcronym, definite start and end point of drawing pattern are hardly to get, so we use another method instead.

\paragraph{Relationship between optimized trajectory and real pattern} To generate candidate patterns, we need understand the relationship between the optimized trajectory and its real pattern, so we designed three experiments to do that. \emph{Experiment 1: Relationship of LSL between trajectory and real pattern} Due to different angles of taping, no matter using hand-hold devices or surveillance cameras, the LSL will deform inevitably. In this experiment, we calculate the relationship of LSL between optimized trajectory and its real pattern. We define the LSL of real pattern as follows: the distance between adjacent dots, no matter horizontally or vertically but not 45-degree angle, is 1. Under this definition, we can get all possible distances between two points: Distance-Set = $\{$ $1, \sqrt{2}, 2, \sqrt{5}, 2 \sqrt{2}$ $\}$, which correspond to $OA, OC, OA', OB, OC'$ shown in Figure.~\ref{fig:all_possible_angles} respectively. Then we record every LSL in trajectory, normalise them by standard length, and use Gaussian kernel function\cite{Gaussian_kernel_function} (see Equation.~\ref{equ:gaussian_kernel_function}) to calculate the density distribution of each type of LSL. The experimental results are shown in Figure.~\ref{fig:distance_dentisity}. 
\begin{equation}
    {k{ \left( {x,x\text{'}} \right) }=\text e\mathop{{}}\nolimits^{{-\frac{{{ \left\Vert {x-x\text{'}} \right\Vert }\mathop{{}}\nolimits^{{2}}}}{{2 \sigma \mathop{{}}\nolimits^{{2}}}}}}}
\label{equ:gaussian_kernel_function}
\end{equation}

\emph{Experiment 2: Relationship of angles between trajectory and real pattern} In this experiment, we try to find the relationship between trajectory and real pattern. As Experiment 1, we first define angle: the smaller angle which two intersectant line segments form. Under this definition, we know that every angle we talk about is less than 180$^{\circ}$. Furthermore, we give all possible angles: Angle-Set = $\{$ 18$^{\circ}$, 27$^{\circ}$, 37$^{\circ}$, 45$^{\circ}$, 53$^{\circ}$, 63$^{\circ}$, 72$^{\circ}$, 90$^{\circ}$, 117$^{\circ}$, 135$^{\circ}$ $\}$ which correspond to $\angle GOH$, $\angle HOI$, $\angle FOH$, $\angle GOI$, $\angle DOF$, $\angle DOI$, $\angle GOH$, $\angle EOI$, $\angle GOH$, $\angle GOH$ shown in Figure.~\ref{fig:all_possible_angles} respectively. We also use Gaussian kernel function to to calculate the density distribution of each type of angle, and the results are shown in Figure.~\ref{fig:angle_dentisity}. 

Intuitively, the more flat the distribution is, the larger variance the actual value will be. If the distribution variance of a feature is small and the mean is close to the standard value, it means that this feature is more effective for our algorithm. The results show significant difference between each distribution of edge length or angle in possible pattern lock and the mean value of each distribution is very close to standard value.

\emph{Experiment 3: Correlation of LSL and angle} Now we have two features that be used to transform optimized trajectory to candidate patterns. So in this experiment, we try to find if we can use one feature to substitute the other one. We introduce Kendall\cite{Kendallcoefficientofconcordance} (see Equation.~\ref{equ:kendall}) and Spearman\cite{Spearmanrankcorrelationcoefficient} (see Equation.~\ref{equ:spearman}) method to calculate the correlation between these two features, i.e. LSL and angle. Furthermore, we also use these two methods to calculate the correlation between LSL of trajectory and its real pattern and angle as well, as a supplement of why LSL and angle detected from video can be used to generate candidate patterns. As Table.~\ref{tab:correlation} shows, we can clearly see that under the Kendall coefficient and Spearman coefficient, both line segment length feature and angle feature have a strong correlation with their corresponding standard values, but there is little correlation between them. In other words, LSL and angle are important features for generating candidate patterns and because of their little correlation, they cannot replace each other.

\begin{equation}
R = \frac{4P}{n(n-1)}-1
\label{equ:kendall}
\end{equation}

\begin{equation}
    \rho = 1- \frac{6\sum
    {d_i^2}}{N(N^2-1)}
\label{equ:spearman}
\end{equation}

\begin{table}[h]
\centering
\begin{tabular}{|c|c|c|}
\hline
\diagbox{Factor}{Correlation}{Algorithm}&Kendall&Spearman\\ 
\hline
Length          & 0.66859106 & 0.81089916 \\ \hline
Angle           & 0.9052319  & 0.9785986  \\ \hline
Length\&Angle   & -0.0596755 & -0.0868365 \\ \hline
\end{tabular}
\label{tab:correlation}
\caption{The correlation of distance, angle and the correlation between them}
\end{table}

\paragraph{Novel method to generate candidate patterns}
We proposed a novel method to generate candidate patterns based on the information we have: a trajectory with some 
redundancy, a sequence of turning points, and the knowledge we mentioned before. In our method, first, we treat each three sequential turning points as a unit, and get the all possible patterns that matches the shape these three turning points form. Then, we use a three point sized window to traverse the turning points, each three points are denoted as a pair of two vectors. For example, from three points \{$(x_1,y_1), (x_2,y_2), (x_3,y_3)$\}, we get a combination of three two-dimensional vectors \{$\vec a = (x_2-x_1,y_2-y_1), \vec b = (x_3-x_2,y_3-y_2), \vec c = (||\vec a||,||\vec b||) $\}.

 
There are 504 possibilities for a three-digit cipher total. Some ciphers like those shown in Figure.~\ref{} may pass four or five keys. Standard vector pairs responding to ciphers are generated, according to the layout of the pattern lock. We calculate the similarity between a standard unit $(\vec u,\vec v, \vec w)$ and a unit$(\vec a,\vec b, \vec c)$ extract from trajectory depending on the cosine similarity of vectors.
\begin{figure}
    \centering
    \begin{minipage}[b]{0.5\textwidth}
	    \subfigure[Vector Units Passed Four Dots]{
			\includegraphics[width=0.22\textwidth]{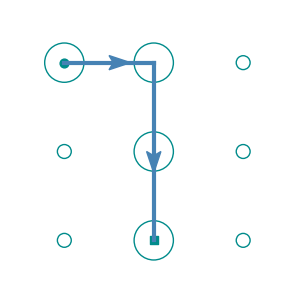} 
			\includegraphics[width=0.22\textwidth]{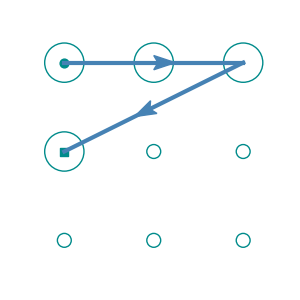}
		\label{fig:valid_patterns}
	}
    \subfigure[Vector Units Passed Five Dots]{
  		 	\includegraphics[width=0.22\textwidth]{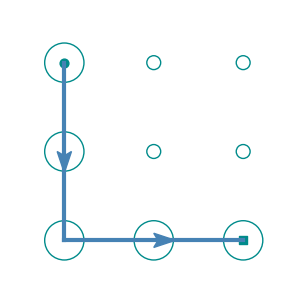}
		 	\includegraphics[width=0.22\textwidth]{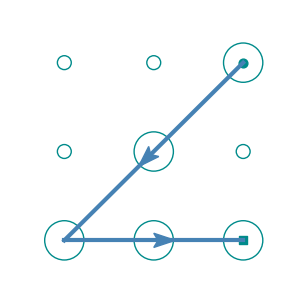}
		\label{fig:invalid_patterns}
    }
	\caption{Some vector units examples that pass more than three dots.}
	\label{fig:three_dots_examples}
	\end{minipage}
\end{figure}

\begin{equation}
\begin{aligned}
S 
&= \frac{1}{2} \times \left(\frac{\vec u \cdot \vec a}{||\vec u||\times||\vec a||}+\frac{\vec v \cdot \vec b}{||\vec v||\times||\vec b||}\right)
\times \theta\\
&+\frac{\vec w \cdot \vec c}{||\vec w||\times||\vec c||} \times (1-\theta)  \qquad (0 \leq \theta \leq 1)
\end{aligned}
\label{formula:similarity}
\end{equation}

To reduce computational complexity, for any vector in unit, if the cosine similarity to the corresponding vector in cipher is less than 0, we consider this cipher impossible and abandon it. The vector $\vec c$ is mainly concerned with the length relation between $\vec a$ and $\vec b$, so a weighting factor $\theta$ is added when dealing with its contribution to similarity. In our initial test, the length relation is not a good basis for similarity because the length can be easily influenced by the condition of the photograph and users' subjective consciousness. And the ratio of length varies little in different cipher, so the weighting factor $\theta$ is set to 0.9.

For each unit, we compute its similarity to max to 504 possible ciphers, some impossible dots(any of the three terms is less than 0) are dropped. Then the possible patterns are reconstructed through combining units together. Because the head and tail of the trajectory may not be in pattern drawing, the process starts from the unit in the middle and goes to both sides. At the beginning, the possible patterns for the mid unit are generated and stored with their confidence(similarity times weight) in a set $P$. When a new unit comes, the set will be updated. The Algorithm.~\ref{code:update_patterns} described the updating of the set. After going through all the units, possible patterns are sorted from high to low according to confidence.


\paragraph{Improved schemes} In order to improve the success rate of restoring the password with fewer attempts, two improved schemes are applied.  

First, we proposed a consensus algorithm to exclude the wrong candidate pattern. As the trajectory deformation of trajectory caused by our camera angle will not affect whether the edges intersect with each other, the intersection between edges of candidate pattern should match the intersection between edges of trajectory edges. As described in Algorithm.~\ref{alogrithm:consistency}, $L_S[]$ is the edge array of line segments.$P[]$ is the set of possible patterns. $tDict$ is the intersection dict of trajectory and $pDict$ is the intersection dict of candidate pattern. Variable $d$ is the difference of indexes in edge array. First, we parse the intersection dict of trajectory and all of the pattern in candidate pattern set. If $pDict$ matches $tDict$, the pattern will be kept, otherwise it will be discarded. The way we parse intersection dict is summerized in Algorithm.~\ref{algorithm:parse_intersection}. variable $d$ is also the difference of indexes in edge array, $c$ is the string that record intersection. For each $d$, if $P[i]$ is intersected with $P[i+d]$, $c$ will append $'T'$, otherwise it will append $'F'$. In the end, for each $d$, we get intersect string $c$. 

\begin{algorithm}[h]
  \caption{Consistency Verification}
  \begin{algorithmic}[1]
    \Require
     $Ls[]$:The set of line segments(edge array);
     
     $P[]$: The set of possible patterns;
     
    \Ensure:
     $N[]$: A new set of possible patterns;
    
    \State $tDict \gets parseIntersect(LS[])$;
    \For{each $pattern \in P_0[]$}
        \State $Lt \gets parsePatternToLineSegment(pattern)$
        \State $pDict \gets parseIntersect(Lt)$
        \State $match = True$
        \For {$d \in [2, 3, 4, 5, 6, 7]$}
            \If {$pDict[d] \notin tDict[d]$}
                \State $match = False$
            \EndIf
        \If{$match$}
            $N.append(pattern)$
        \EndIf
        \EndFor
      
    \EndFor
    \State\Return $N$;
  \end{algorithmic}
 \label{alogrithm:consistency}
\end{algorithm}

\begin{algorithm}[h]
  \caption{Parse Intersection}
  \begin{algorithmic}[1]
    \Require
        $P[]$: The set of line segments(edge array);
    \Ensure:
        $intersectDict{}$: A dict of intersect
    \For {$d \in [2,3,4,5,6,7]$}
        \State $String ~ c \gets CountIntersect(d)$
        \State $intersectDict(d) = c$
    \EndFor
    \State\Return $intersectDict$;
  \end{algorithmic}
 \label{algorithm:parse_intersection}
\end{algorithm}

As Mentioned in Subsection.~\ref{subsec:track_hand_motion}, we also tracked other hand key points and get trajectories to generate candidate patterns. For the candidate patterns generated by different trajectories, the confidence of each pattern is accumulated, and finally the candidate patterns are sorted to get a final results. By applying these two schemes, we can successfully crack the lock pattern in fewer attempts, the comparison results of whether to use the two schemes are presented in Section.~\ref{subsec:effects_setting}.

\begin{figure}
	\centering
    		\begin{minipage}[b]{0.5\textwidth}
   		 	\includegraphics[width=0.3\textwidth]{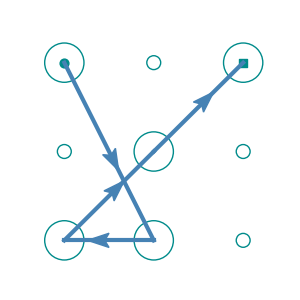}
		 	\includegraphics[width=0.3\textwidth]{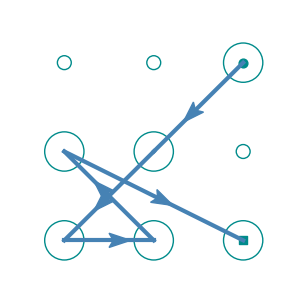}
		 	\includegraphics[width=0.3\textwidth]{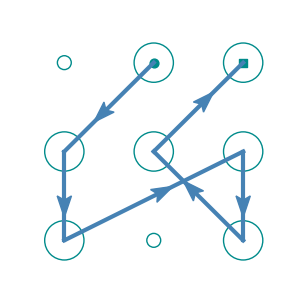}
    		\end{minipage}
	\label{fig:cross_patterns}
	\caption{Some patterns that contain intersections.}
\end{figure}


\section{Evaluation}
\label{sec:evaluation}


In this section, we demonstrate the simulation experiment and the results. The experimental results reported in this section are based on data collected from the authors only. We plan to recruite a number of human participants to validate the performance of the proposed attacking pipeline using more realistic data from real world users.
Based on these test samples, we can guess over 90\% of videos filmed in scenario 1 and 60\% in scenario 2 in ten attempts. 



\subsection{Experiment Setup}
\label{subsec:environment_setup}

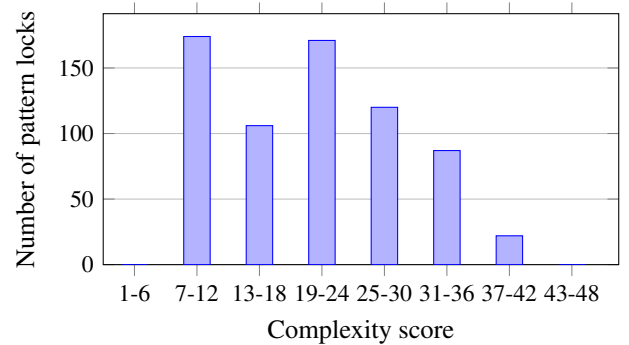
\begin{figure}[htbp]
    \centerline{
    \pgfplotsset{width=24em,height = 14em}
    \pgfplotstableread[col sep=comma]{data/pattern_complexity.csv}\datacsv
    \begin{tikzpicture}
    \begin{axis}[
    ybar,
    xlabel=Complexity score,
    ylabel=Number of pattern locks,
    ymin = 0,
    xmin = 0,
    xtick = data,
    xticklabels ={1-6, 7-12, 13-18, 19-24, 25-30, 31-36, 37-42, 43-48 },
    x tick label style={font=\small},
    y tick label style={font=\small},
    ymajorgrids=true,
    ylabel near ticks,
    xlabel near ticks,
    ]
    \addplot table[x=a, y=b] {\datacsv};
    \end{axis}
    \end{tikzpicture}
    }
	\caption{The distribution of pattern complexity scores}
	\label{fig:pattern complexity}
\end{figure}

\paragraph{Pattern selection}
TBD

\paragraph{Device selection}
TBD

\paragraph{Attack scenarios}

As mentioned in [cite, threat model], our system focusing on working in 2 main scenarios, so the detailed information is as follows:

\textbf{1}.Surveillance cameras:To simulate surveillance cameras, we use tripod to fix video-taping devices with 2.5 meters high from ground. The horizontal distances from taping devices are 1 meter, 3 meters, 5 meters, and direction includes in front , and left to the volunteer.All volunteers has the same position of drawing patterns of sitting near a table with target phone laid on it.

\textbf{2}.Face-to-face:To simulate this scenario, the video-taping device is held by another volunteer and is at the same vertical height as the target device.The horizontal distances between these two devices differ from 2 meter, 3 meters, 5 meters and 10 meters.Target device is held by a volunteer, with him sitting near a table.

To summarize, we in total record xx videos with xx different patterns.

\subsection{Overall Success Rate}
In our experiment, over 99.3\% of video samples in scenario 1 can be successfully cracked in 20 attempts. In scenario 2, we can crack over 92\% of video samples in 20 attempts. The relation between attempts times and success rate is shown in figure 10.

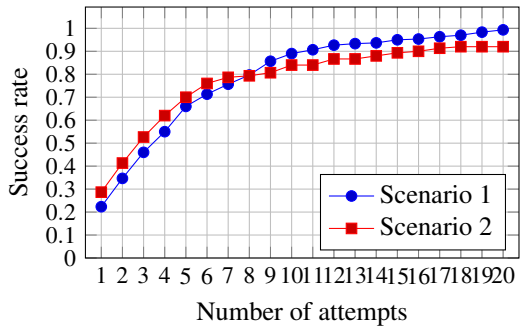
\begin{figure}[htbp]
    \centerline{
    \pgfplotsset{width=14em,height = 14em}
    \begin{tikzpicture}
    \begin{axis}[
    x=8pt, enlarge x limits={true, abs value=0.75},
    ymin=0, enlarge y limits={upper, value=0.1},
    grid=major,
    xlabel=Number of attempts,
    ylabel=Success rate,
    ylabel near ticks,
    xlabel near ticks,
    xtick = data,
    ytick distance=0.1,
    x tick label style={font=\small},
    y tick label style={font=\small},
    legend pos=south east,
    ]
    \addplot coordinates{                                
(1,0.22333333333333333)
(2,0.3466666666666667)
(3,0.46)
(4,0.55)
(5,0.66)
(6,0.7133333333333334)
(7,0.7566666666666667)
(8,0.7966666666666666)
(9,0.8566666666666667)
(10,0.89)
(11,0.9066666666666666)
(12,0.9266666666666666)
(13,0.9333333333333333)
(14,0.9366666666666666)
(15,0.95)
(16,0.9533333333333334)
(17,0.9633333333333334)
(18,0.97)
(19,0.9833333333333333)
(20,0.9933333333333333)};
 \addplot coordinates{                                
(1,0.2866666666666667)
(2,0.41333333333333333)
(3,0.5266666666666666)
(4,0.62)
(5,0.7)
(6,0.76)
(7,0.7866666666666666)
(8,0.7933333333333333)
(9,0.8066666666666666)
(10,0.84)
(11,0.84)
(12,0.8666666666666667)
(13,0.8666666666666667)
(14,0.88)
(15,0.8933333333333333)
(16,0.9)
(17,0.9133333333333333)
(18,0.92)
(19,0.92)
(20,0.92)};
\legend{Scenario 1, Scenario 2}
    \end{axis}
    \end{tikzpicture}
	}
	\caption{The successful cracking rate in different number of attempts.}
	\label{fig:overall_success}
\end{figure}

TBD

\subsection{Impact of Distance}
In both scenarios, the vertical distance between the camera and the target phone is relatively single. We mainly consider the impact of horizontal distance on success rate. There are four varieties in distance: 1m, 2m, 3m and 5m. 

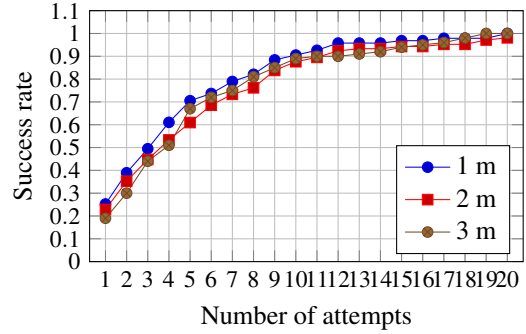
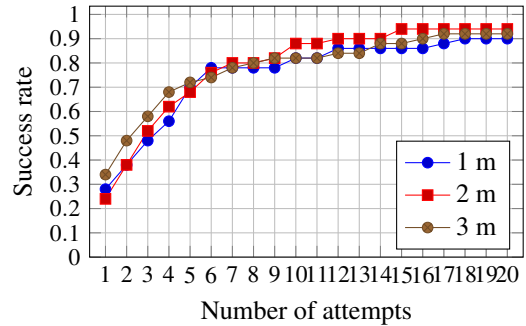
\begin{figure}[htbp]
    \subfigure[Scenario 1]{
    \pgfplotsset{width=14em,height = 14em}
    \begin{tikzpicture}
    \begin{axis}[
    x=8pt, enlarge x limits={true, abs value=0.75},
    ymin=0, enlarge y limits={upper, value=0.1},
    grid=major,
    xlabel=Number of attempts,
    ylabel=Success rate,
    ylabel near ticks,
    xlabel near ticks,
    xtick = data,
    ytick distance=0.1,
    x tick label style={font=\small},
    y tick label style={font=\small},
    legend pos=south east,
    ]
    \addplot coordinates{                                
(1,0.25263157894736843)
(2,0.3894736842105263)
(3,0.49473684210526314)
(4,0.6105263157894737)
(5,0.7052631578947368)
(6,0.7368421052631579)
(7,0.7894736842105263)
(8,0.8210526315789474)
(9,0.8842105263157894)
(10,0.9052631578947369)
(11,0.9263157894736842)
(12,0.9578947368421052)
(13,0.9578947368421052)
(14,0.9578947368421052)
(15,0.968421052631579)
(16,0.968421052631579)
(17,0.9789473684210527)
(18,0.9789473684210527)
(19,0.9789473684210527)
(20,1.0)
};
 \addplot coordinates{                                
(1,0.22857142857142856)
(2,0.3523809523809524)
(3,0.44761904761904764)
(4,0.5333333333333333)
(5,0.6095238095238096)
(6,0.6857142857142857)
(7,0.7333333333333333)
(8,0.7619047619047619)
(9,0.8380952380952381)
(10,0.8761904761904762)
(11,0.8952380952380953)
(12,0.9238095238095239)
(13,0.9333333333333333)
(14,0.9333333333333333)
(15,0.9428571428571428)
(16,0.9428571428571428)
(17,0.9523809523809523)
(18,0.9523809523809523)
(19,0.9714285714285714)
(20,0.9809523809523809)
};
 \addplot coordinates{                                
(1,0.19)
(2,0.3)
(3,0.44)
(4,0.51)
(5,0.67)
(6,0.72)
(7,0.75)
(8,0.81)
(9,0.85)
(10,0.89)
(11,0.9)
(12,0.9)
(13,0.91)
(14,0.92)
(15,0.94)
(16,0.95)
(17,0.96)
(18,0.98)
(19,1.0)
(20,1.0)
};
\legend{1 m, 2 m, 3 m}
    \end{axis}
    \end{tikzpicture}
}
	\subfigure[Scenario 2]{
    \pgfplotsset{width=14em,height = 14em}
    \begin{tikzpicture}
    \begin{axis}[
    x=8pt, enlarge x limits={true, abs value=0.75},
    ymin=0, enlarge y limits={upper, value=0.1},
    grid=major,
    xlabel=Number of attempts,
    ylabel=Success rate,
    ylabel near ticks,
    xlabel near ticks,
    xtick = data,
    ytick distance=0.1,
    x tick label style={font=\small},
    y tick label style={font=\small},
    legend pos=south east,
    ]
    \addplot coordinates{                                
(1,0.28)
(2,0.38)
(3,0.48)
(4,0.56)
(5,0.7)
(6,0.78)
(7,0.78)
(8,0.78)
(9,0.78)
(10,0.82)
(11,0.82)
(12,0.86)
(13,0.86)
(14,0.86)
(15,0.86)
(16,0.86)
(17,0.88)
(18,0.9)
(19,0.9)
(20,0.9)
};
 \addplot coordinates{                                
(1,0.24)
(2,0.38)
(3,0.52)
(4,0.62)
(5,0.68)
(6,0.76)
(7,0.8)
(8,0.8)
(9,0.82)
(10,0.88)
(11,0.88)
(12,0.9)
(13,0.9)
(14,0.9)
(15,0.94)
(16,0.94)
(17,0.94)
(18,0.94)
(19,0.94)
(20,0.94)
};
\addplot coordinates{                                
(1,0.34)
(2,0.48)
(3,0.58)
(4,0.68)
(5,0.72)
(6,0.74)
(7,0.78)
(8,0.8)
(9,0.82)
(10,0.82)
(11,0.82)
(12,0.84)
(13,0.84)
(14,0.88)
(15,0.88)
(16,0.9)
(17,0.92)
(18,0.92)
(19,0.92)
(20,0.92)
};
\legend{1 m, 2 m, 3 m}
    \end{axis}
    \end{tikzpicture}
	}
	\caption{The successful rate in different shooting distance.}
	\label{fig:distance_rate}
\end{figure}

As shown in figure \ref{fig:distance_rate}, for samples filmed by monitor, we get the highest success rate when the horizontal distance is 1m. As the distance increases, the success rate goes down both in five attempts and 20 attempts. In face-to-face shooting scenario, the success rate is not affected by distance.

\subsection{Impact of Phone Posture}
In scenario 1, as mentioned in \ref{subsec:new_challenges}, the orientation of the phone can be identified by detecting the orientation of the human face. There is no need for \ModelAcronym to transfer the phone in video into completely front view. With knowing the head orientation of the phone, even the orientation identified is somewhat different from the real orientation, \ModelAcronym can successfully resort the pattern lock. 


\begin{figure}[htbp]
    \centerline{
    \pgfplotsset{width=14em,height = 14em}
    \begin{tikzpicture}
    \begin{axis}[
    ybar,
    bar width=3pt,
    x=40pt, enlarge x limits={true, abs value=0.75},
    ymin=0, enlarge y limits={upper, value=0.1},
    ymajorgrids=true,
    xlabel= Directions,
    ylabel=Success rate,
    ylabel near ticks,
    xlabel near ticks,
    xticklabels ={main direction, turn left, turn right},
    xtick = data,
    ytick distance=0.1,
    x tick label style={font=\small},
    y tick label style={font=\small},
    every axis legend/.append style={
    legend columns=1,font=\small},
    legend pos=outer north east,
    ]
\addplot coordinates{                                
 (1,0.7) (2,0.65) (3,0.55) };
\addplot coordinates{                                
 (1,0.8) (2,0.75) (3,0.65) };
 \legend{Front Filming, Left Filming};
    \end{axis}
    \end{tikzpicture}
	}
	\caption{Impact of phone grip angle.}
	\label{fig:angle_rate}
\end{figure}
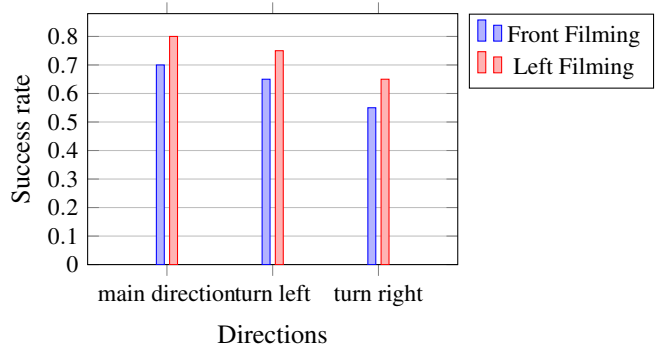

\begin{figure}[htbp]
    \centerline{
    \pgfplotsset{width=14em,height = 14em}
    \begin{tikzpicture}
    \begin{axis}[
    x=8pt, enlarge x limits={true, abs value=0.75},
    ymin=0, enlarge y limits={upper, value=0.1},
    grid=major,
    xlabel=Number of attempts,
    ylabel=Success rate,
    ylabel near ticks,
    xlabel near ticks,
    xtick = data,
    ytick distance=0.1,
    x tick label style={font=\small},
    y tick label style={font=\small},
    legend pos=south east,
    ]
    \addplot coordinates{                                
(1,0.15) 
(2,0.2)
(3,0.25)
(4,0.275)
(5,0.43)
(10,0.63)
(15,0.69)
(20,0.7416)};
 \addplot coordinates{                                
 (1,0.25) (2,0.3
) (3,0.4
)
 (4,0.5
) (5,0.583
) (10,0.69
)
 (15,0.74
) (20,0.825
)};
\addplot coordinates{                                
 (1,0.2) (2,0.25
) (3,0.3
)
 (4,0.4
) (5,0.5
) (10,0.6
)
 (15,0.7
) (20,0.8
)};
\legend{0 degrees, 0-5 degrees,>5 degrees}
    \end{axis}
    \end{tikzpicture}
	}
	\caption{The successful cracking rate in different number of attempts.}
	\label{fig:rate_grip_angle}
\end{figure}
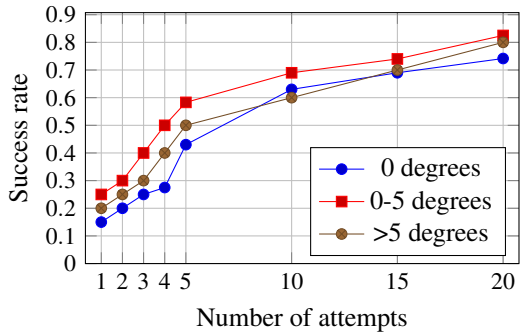

In scenario 2,the phone's screen is projected onto  the vertical plane at one angle, depending on how the user holds it. In the experiment, we found that it was difficult for the user to fix the position of holding the phone while entering pattern, and it was difficult to determine the angle between the phone and the vertical direction. As we track the visible part of hand, which is not in direct contact with the phone screen in scenario 2, the Angle of the phone's grip does not directly affect the success rate. Nonetheless, the impact of this Angle on the success rate cannot be ignored. We compared the success rate by dividing the samples into three categories according to the degree of skew. The results
is show on Figure.~\ref{fig:rate_grip_angle}.


\subsection{Impact of Pattern Type}

\begin{figure}[htbp]
    \centerline{
    \pgfplotsset{width=14em,height = 14em}
    \begin{tikzpicture}
    \begin{axis}[
    x=8pt, enlarge x limits={true, abs value=0.75},
    ymin=0, enlarge y limits={upper, value=0.1},
    grid=major,
    xlabel=Number of attempts,
    ylabel=Success rate,
    ylabel near ticks,
    xlabel near ticks,
    xtick = data,
    ytick distance=0.1,
    legend pos=south east,
    ]
    \addplot coordinates{    
(1,0.16521739130434782)
(2,0.2956521739130435)
(3,0.3826086956521739)
(4,0.4782608695652174)
(5,0.6086956521739131)
(6,0.6521739130434783)
(7,0.7130434782608696)
(8,0.7652173913043478)
(9,0.8347826086956521)
(10,0.8347826086956521)
(11,0.8695652173913043)
(12,0.8869565217391304)
(13,0.9043478260869565)
(14,0.9043478260869565)
(15,0.9130434782608695)
(16,0.9130434782608695)
(17,0.9391304347826087)
(18,0.9478260869565217)
(19,0.9739130434782609)
(20,0.9826086956521739)};
    \addplot coordinates{                              
(1,0.2288135593220339)
(2,0.3305084745762712)
(3,0.4661016949152542)
(4,0.5508474576271186)
(5,0.6610169491525424)
(6,0.711864406779661)
(7,0.7542372881355932)
(8,0.788135593220339)
(9,0.864406779661017)
(10,0.923728813559322)
(11,0.9322033898305084)
(12,0.9491525423728814)
(13,0.9491525423728814)
(14,0.9576271186440678)
(15,0.9661016949152542)
(16,0.9745762711864406)
(17,0.9745762711864406)
(18,0.9830508474576272)
(19,0.9830508474576272)
(20,1.0)};
    \addplot coordinates{                                
(1,0.31343283582089554)
(2,0.4626865671641791)
(3,0.582089552238806)
(4,0.6716417910447762)
(5,0.746268656716418)
(6,0.8208955223880597)
(7,0.835820895522388)
(8,0.8656716417910447)
(9,0.8805970149253731)
(10,0.9253731343283582)
(11,0.9253731343283582)
(12,0.9552238805970149)
(13,0.9552238805970149)
(14,0.9552238805970149)
(15,0.9850746268656716)
(16,0.9850746268656716)
(17,0.9850746268656716)
(18,0.9850746268656716)
(19,1.0)
(20,1.0)};
    \legend{simple, medium, complex}
    \end{axis}
    \end{tikzpicture}
	}
	\caption{The successful cracking rate in different types of patterns.}
	\label{fig:rate_complexity}
\end{figure}
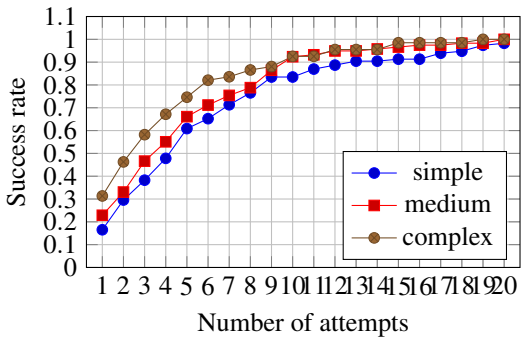

As mentioned in \ref{subsec:environment_setup}, we calculate the complexity score of each pattern. In this experiment we divided all the patterns into three groups: complex, medium and simple. The success rate for each type of pattern is shown on Figure \ref{fig:rate_complexity}.


For the two main factors that affect password complexity: length and cross or not, we also set an experiment to study each factor's impacts on success rate.

\begin{table}[!htbp]
\centering
\begin{tabular}{|c|c|c|}
\hline
\diagbox{factors}{success rate}{factors}&long&short\\ 
\hline
cross&90\%&80\%\\
\hline
no cross&95\%&90\%\\
\hline
\end{tabular}
\caption{The successful cracking rate in different types of patterns}
\label{tab:rate_type}
\end{table}



\subsection{Impact of Settings}
\label{subsec:effects_setting}
There are some adjustable parameters in our system such as the number of frames detected and the key points in hand. In this section, we evaluate the effects on success rate of different settings. 
\paragraph{Frame Detected}
In our approach, once a phone is detected, the following several frames will be used as the input image to detect phone and hand. Then one frame which have the detection results with highest confidence is chosen as the tracking start frame. The number of frames is set to 30 in our overall experiment, but there remains some options. When the number is too small, the detection results may be unsatisfied. When the number is too large, the tracking may start after the drawing of pattern. So we did the experiment in different number settings.
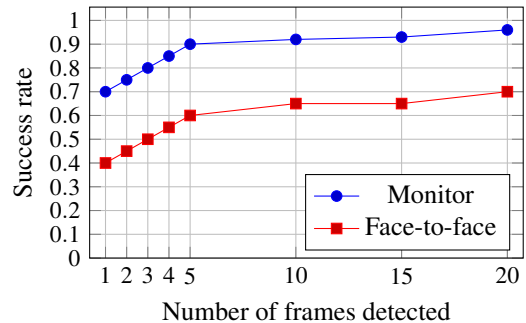
\begin{figure}[htbp]
    \centerline{
    \pgfplotsset{width=14em,height = 14em}
    \begin{tikzpicture}
    \begin{axis}[
    x=8pt, enlarge x limits={true, abs value=0.75},
    ymin=0, enlarge y limits={upper, value=0.1},
    grid=major,
    xlabel=Number of frames detected,
    ylabel=Success rate,
    ylabel near ticks,
    xlabel near ticks,
    xtick = data,
    ytick distance=0.1,
    x tick label style={font=\small},
    y tick label style={font=\small},
    legend pos=south east,
    ]
    \addplot coordinates{                                
 (1,0.7) (2,0.75) (3,0.80)
 (4,0.85) (5,0.90) (10,0.92)
 (15,0.93) (20,0.96)};
 \addplot coordinates{                                
 (1,0.4) (2,0.45) (3,0.50)
 (4,0.55) (5,0.60) (10,0.65)
 (15,0.65) (20,0.70)};
\legend{Monitor,Face-to-face}
    \end{axis}
    \end{tikzpicture}
	}
	\caption{The successful cracking rate in different number of frames detected.}
	\label{fig:cracking_rate_numframes}
\end{figure}
As the figure...

\paragraph{KeyPoints}
In scenario 1, the tip of index finger is the target we chose for tracking because its motion reveals the pattern drawing quite well. There are some other parts in hand that can be treated as targets such as the second and the first joint of index finger. We chose different parts as the tracking target and got their influence on success rate.

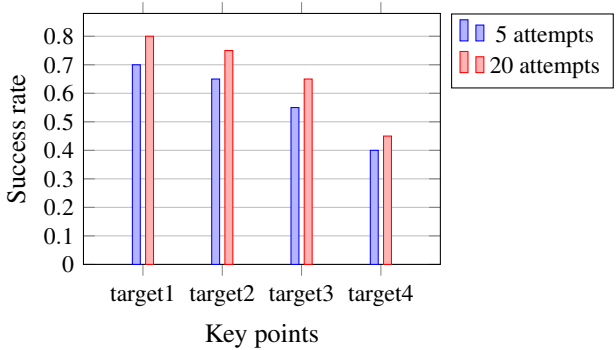
\begin{figure}[htbp]
    \centerline{
    \pgfplotsset{width=14em,height = 14em}
    \begin{tikzpicture}
    \begin{axis}[
    ybar,
    bar width=3pt,
    x=30pt, enlarge x limits={true, abs value=0.75},
    ymin=0, enlarge y limits={upper, value=0.1},
    ymajorgrids=true,
    xlabel= Key points,
    ylabel=Success rate,
    ylabel near ticks,
    xlabel near ticks,
    xticklabels ={target1,target2,target3,target4},
    xtick = data,
    ytick distance=0.1,
    x tick label style={font=\small},
    y tick label style={font=\small},
    every axis legend/.append style={
    legend columns=1,font=\small},
    legend pos=outer north east,
    ]
\addplot coordinates{                                
 (1,0.7) (2,0.65) (3,0.55)
(4,0.40) };
\addplot coordinates{                                
 (1,0.8) (2,0.75) (3,0.65)
(4,0.45) };
 \legend{5 attempts, 20 attempts};
    \end{axis}
    \end{tikzpicture}
	}
	\caption{The successful cracking rate using different key points.}
	\label{fig:cracking_rate_anchors}
\end{figure}


\section{Discussion}
\label{sec:discussion}
\subsection{Limitations}
\label{subsec:limitation}

Our research works well under our assumptions mentioned in Section~\ref{subsec:assumptions}, however, we do have some limitations. First, in some places (e.g. railway station, or airport), surveillance cameras may deployed much higher than our assumption. In that case, drawing hand identification and object tracking algorithm may lose efficacy, which can lead our work become invalid. Second, though our work do not fingertip in the video, but we still need parts of drawing hand continuously appear in the video, for we do not have prediction algorithm for missing key points. Third, there are certain people who has different drawing habit. For example, one may have a very simple pattern as his locking pattern, and he can draw this pattern with only finger motion but his hand barely moves. In that case, our work may lose efficacy.

\subsection{Mitigations}
\label{subsec:mitigation}

Based on our work, we clearly give the threateness of video-attack towards pattern lock. A potential defence of these attack is remembering cover hand when people drawing pattern, or input text-based password, no matter in public place or in a small room. Losing visual of keypoints always has big effect on these attacks. Another possible defence is that, people can use multiple authentication methods to protect his personal information. In recent years, authentication based on biological characteristics such as fingerprints\cite{maltoni2009handbook}, iris\cite{daugman2009iris} and face\cite{ijiri2006security} recognition are widely used. These bio-based authentication may vulnerable by other attacks, but are relatively secure to video-based attack. So a better way to protect personal information may use multi-factor authentication.


\section{Conclusion}

In this paper, we proposed a automatic video-based attack towards pattern lock. Our Experiments showed that, although some prior work gave some evidences that pattern lock is vulnerable, but the actual threaten is still underestimated. An experienced adversary can utilize different ways, such as face-to-face taping or hacking surveillance cameras, to guess target user's pattern in very short time, and high success rate.  


\bibliographystyle{plain}
\bibliography{ref}

\newpage
\section{Appendix}

\begin{algorithm}[htbp]
  \caption{Check}
  \begin{algorithmic}[1]
    \Require
        $trajectory$: The temporary trajectory while tracking;
        $CR$: The threshold of number of frames to check moving;
    \Ensure:
        $State$: The temporary trajectory is valid or not.
    \State $AD = getAvgDistance(trajectory)$
    \State $State = True$
    \State $d \gets distance(trajectory[-CR:])$
    \If{$d\ \leq\ 5$}
        \State $countStatic += 1$
    \Else
    \State $countStatic = 0$
    \EndIf
    \If{$countStatic \geq 20$}
        \State $State = False$
    \EndIf
    \If{$distance(trajectory[-2:])\geq 2 \times\ AD$}
        \State $State=False$
    \EndIf
    \State \Return $State$
  \end{algorithmic}
\label{code:check}
\end{algorithm}

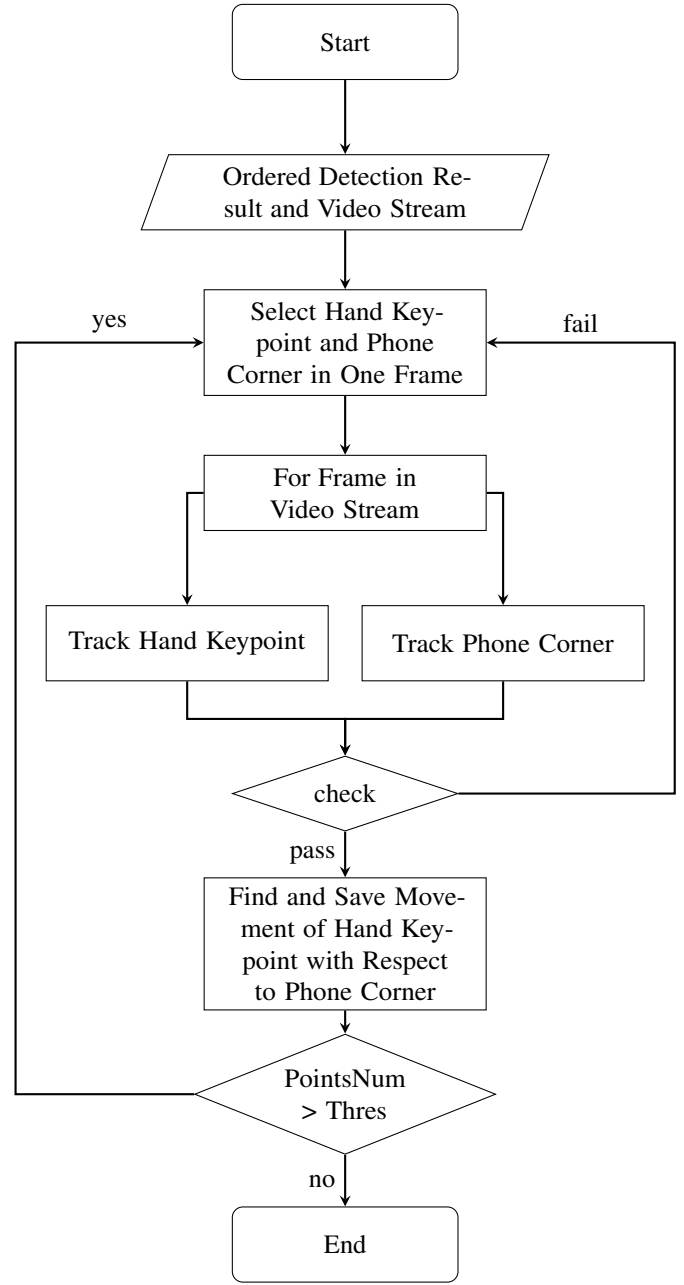
\begin{figure}[htbp]
\usetikzlibrary{calc}
\tikzstyle{startstop} = [rectangle, rounded corners, minimum width=3cm, minimum height=1cm, text centered,draw=black]
\tikzstyle{io}        = [trapezium, trapezium left angle=70, trapezium right angle=110, minimum width=3cm, inner xsep = -15pt, minimum height=1cm, text centered, draw=black]
\tikzstyle{process}   = [rectangle, minimum width=3cm, minimum height=1cm, inner ysep=0, text centered, draw=black, text width = 10em]
\tikzstyle{decision}  = [diamond,shape aspect=2.5, minimum width=3cm, minimum height=1cm, inner xsep=0,text centered, draw=black]
\tikzstyle{arrow}     = [thick,->,>=stealth]

\centering
\begin{tikzpicture}[node distance = 2cm]
  \node (start) [startstop] {Start};
  \node (in1) [io, below of = start, text width = 15em] {Ordered Detection Result and Video Stream}; 
  \node (pro1) [process, below of=in1, minimum height=4em] {Select Hand Keypoint and Phone Corner in One Frame};
  \node (pro2) [process, below of=pro1] {For Frame in Video Stream};
  \node (dec1) [decision, below of=pro2,  yshift = -2cm] {check};
  \node (pro3) [process, below of =pro2, xshift=-2.1cm] {Track Hand Keypoint};
  \node (pro4) [process, below of=pro2, xshift=2.1cm] {Track Phone Corner};
  \node (pro5) [process, below of=dec1, minimum height=5em] {Find and Save Movement of Hand Keypoint with Respect to Phone Corner};
  \node (pro6)  [decision, below of=pro5, text width = 5em] {PointsNum > Thres};
  \node (stop) [startstop, below of=pro6] {End};
  \draw[arrow] (start)  -- (in1);
  \draw[arrow] (in1) -- (pro1);
  \draw[arrow] (pro1)-- (pro2);
  \draw[arrow] (pro2.east) -|  (pro4);
  \draw[arrow] (pro2.west) -|  (pro3);
  \draw[arrow] (pro3) --($(pro3.south)-(0,0.5)$) -| (dec1);
  \draw[arrow] (pro4) --($(pro4.south)-(0,0.5)$) -| (dec1);
  \draw[arrow] (dec1) -- node[anchor=east] {pass} (pro5);
  \draw[arrow] (dec1.east) -| ($(pro1.east) + (2.5, 0)$) --node[anchor=south] {fail} (pro1.east);
  \draw[arrow] (pro5) -- (pro6);
  \draw[arrow] (pro6.west) -| ($(pro1.west) - (2.5, 0)$)-- node[anchor=south] {yes} (pro1.west);
  \draw[arrow] (pro6) -- node[anchor=east] {no} (stop);
\end{tikzpicture}
\caption{The flowchart}\label{fig:trackflowchart}
\end{figure}

\begin{algorithm*}[htbp]
  \caption{ Update Possible Patterns Set}
  \begin{algorithmic}[1]
    \Require
     $LS[]$:The set of line segments;
     
     $P_0[]$: The old set of possible patterns;
     
     $U<line1,line2>$: A incoming unit;
     
     $C[]$: The set of possible ciphers of U;
     
    \Ensure:
     $P_1[]$: A new set of possible patterns;
    \For{each $pattern \in P_0[]$}
        \State $nextUnit \gets getNext(LS[],pattern)$;
        \State $lastUnit \gets getLast(LS[],pattern)$;
      \If{U == nextUnit}
        \State $overlapSegment \gets getOverlap(pattern.passed,U)$;
      \For{each cipher in C}
        \If{cipher[overlapSegment] == pattern[overlapSegment]}
        \State $passedUnit \gets  pattern.passedUnit \cup U.line2$;
        \State $keys \gets pattern.keys \cup cipher.lastkey $;
        \State $confidence \gets  pattern.confidence + dots.similarity \times U.weight$;
        \State $newpattern \gets  <keys,passedUnit,confidence>$;
        \EndIf
        \EndFor
        \EndIf
    \If{U == lastUnit}
        \State $overlapSegment \gets getOverlap(pattern.passed,U)$;
      \For{each cipher in C}
        \If{cipher[overlapSegment] == pattern[overlapSegment]}
        \State $passedUnit \gets  pattern.passedUnit \cup U.line2l$;
        \State $keys \gets pattern.keys \cup cipher.lastkey $;
        \State $confidence \gets  pattern.confidence + dots.similarity \times U.weight$;
        \State $newpattern \gets  <keys,passedUnit,confidence>$;
        \EndIf
        \EndFor
      \EndIf
      \State $P_0 \gets P_0\cup newpattern$;
      \
    \EndFor
    \State $P_1 \gets P_0$;
    \Return $P_1$;
  \end{algorithmic}
 \label{code:update_patterns}
\end{algorithm*}

\begin{algorithm}[htbp]
  \caption{Process}
  \begin{algorithmic}[1]
    \Require
     $startNum$:The start frame number;
     
     $frames[]$: The frames of video stream;
     
    \Ensure:
     $trajectory[]$: Point coordinates obtained by tracking;
    \For{each $frameNum, frame \in enumerate(frames)$}
        \If{$frameNum < startNum$}
            \State $continue$;
        \EndIf
        \State $cornerBox \gets detectCorner(frame)$;
        \State $fingerBoxs \gets detectFinger(frame)$;
        \State $
        \begin{aligned}
        trajectory \gets track(&cornerBox,fingerBoxs[1], \\ &frames, frameNum)
        \end{aligned}
        $;
        \If{$trajectory\ !=\ None$}
            \State \Return $trajectory$;
        \EndIf
        \State $
        \begin{aligned}
        trajectory \gets track(&cornerBox,fingerBoxs[1], \\ &frames, frameNum)
        \end{aligned}
        $;
        \If{$trajectory\ !=\ None$}
            \State \Return $trajectory$;
        \EndIf
    \EndFor
    \State \Return $[]$;
    
  \end{algorithmic}
\label{code:process}
\end{algorithm}

\begin{algorithm}[htbp]
  \caption{track}
  \begin{algorithmic}[1]
    \Require
        $startNum$:The start frame number;
     
        $frames[]$: The frames of video stream;
     
        $cBox[]$ The location of phone corner;
        
        $pBox[]$ The location of phoneBox;
     
        $fBox[]$ The location of finger box;
    \Ensure:
        $trajectory[]$: Point coordinates obtained by tracking;
    \State $trajectory = []$
    \State $inside = False$
    \For{$frameNum, frame \in enumerate(frames)$}
        \If{$frameNum < startNum$}
            \State $continue$
        \EndIf
        \If{$frameNum == startNum$}
            \State $fingerTracker.init(fBox)$
            \State $cornerTracker.init(cBox)$
        \EndIf
        \State $fCenter \gets Center(fBox)$
        \State $cCenter \gets Center(cBox)$
        \State $
        \begin{aligned}
        trajectory.append(&fCenter[0]-cCenter[0], \\
        &fCenter[1]-cCenter[1])
        \end{aligned}
        $;
        \State $ok \gets check(trajectory)$
        \If{ $!\ ok$}: 
            \State \Return $[]$
        \EndIf
        \State $
        \begin{aligned}
        fingerOK, fBox \gets 
        fingerTracker.update(frame)
        \end{aligned}
        $
        \State $
        \begin{aligned}
        cornerOK,cBox \gets 
        cornerTracker.update(frame)
        \end{aligned}
        $
        \If{$ !\ fingerOK \ || \ !\ cornerOK$}
        \State \Return $[]$;
        \EndIf
        \If{$!\ inside(fBox, pBox)$}:
            \If {$!\ inside:$}
                    $continue$
            \Else:
                    $break$
            \EndIf
        \Else:
            \State $inside \gets True$;
        \EndIf
    \EndFor
    \State \Return $trajectory$
  \end{algorithmic}
\label{code:track}
\end{algorithm}
\end{document}